\documentclass[a4paper,fleqn]{cas-dc}

\usepackage[numbers]{natbib}

\usepackage{graphicx}
\usepackage{bbm}
\usepackage{amsmath}
\usepackage[subpreambles=true]{standalone}
\usepackage{mathrsfs}
\usepackage{color,soul}
\usepackage{algorithm}
\usepackage{algpseudocode}
\usepackage{hyperref} 
\usepackage{amsmath,amssymb}

\usepackage{amsthm}
\usepackage{subfig}
\newtheorem{theorem}{Theorem}[section]

\def\tsc#1{\csdef{#1}{\textsc{\lowercase{#1}}\xspace}}
\tsc{WGM}
\tsc{QE}
\tsc{EP}
\tsc{PMS}
\tsc{BEC}
\tsc{DE}

\begin{document}
\let\WriteBookmarks\relax
\def\floatpagepagefraction{1}
\def\textpagefraction{.001}
\shorttitle{Energy storage applications for low voltage consumers in Uruguay}
\shortauthors{Md Umar Hashmi et~al.}

\title [mode = title]{Energy storage applications for low voltage consumers in Uruguay}

\author[1,3,6]{Md Umar Hashmi}
\author[2,3,4]{Jos\'e Luis Horta}
\author[2,3]{Diego Kiedanski}
\author[5,6]{Lucas Pereira}
\author[1,3]{Ana Bu\v{s}i\'c}
\author[2,3]{Daniel Kofman}
\address[1]{INRIA, DI ENS, Ecole Normale Sup\'erieure, CNRS, PSL Research University, Paris, France}
\address[2]{T\'el\'ecom ParisTech, France}
\address[3]{Laboratory of Information, Networking and Communication Sciences (LINCs), Paris, France}
\address[4]{ICT4V, Montevideo, Uruguay}
\address[5]{Madeira Interactive Technologies Institute/LARSYS, Funchal and T\'enico Lisboa, Universidade de Lisboa, 1049-001 Lisbon, Portugal}
\address[6]{prsma.com, Funchal, Portugal}
%
%
%
%
%
%
%


\begin{abstract}
	Energy storage can be used for many applications in the Smart Grid such as energy arbitrage, peak demand shaving, power factor correction, energy backup to name a few, and can play a major role at increasing the capacity of power networks to host renewable energy sources. Often, storage control algorithms will need to be \textit{tailored} according to power networks billing structure, reliability restrictions, and other local power networks norms. 
	In this paper we explore residential energy storage applications in Uruguay, one of the global leaders in renewable energies, where new low-voltage consumer contracts were recently introduced.  Based on these billing mechanisms, we focus on energy arbitrage and reactive energy compensation with the aim of minimizing the cost of consumption of an end-user. 
	Given that in the new contacts the buying and selling price of electricity are equal and that reactive power compensation is primarily governed by the installed converter, the storage operation is not sensitive to parameter uncertainties and, therefore, no lookahead is required for decision making. A threshold-based \textit{hierarchical} controller is proposed which decides on the optimal active energy for arbitrage and uses the remaining converter capacity for reactive power compensation, which is shown to increase end-user profit.
	Numerical results indicate that storage could be profitable, even considering battery degradation, under some but not
	all of the studied contracts. For the cases in which it is not, we propose the
	best-suited contract.
	Results presented here can be naturally applied whenever the tariff structure satisfies the hypothesis considered in this work.

\end{abstract}

\begin{keywords}
	Energy storage \sep Electricity billing \sep Energy arbitrage \sep Reactive compensation \sep Net energy metering \sep Hierarchical control
\end{keywords}


\maketitle

\section{Introduction}

Electric power systems are undergoing major transformations because of changes in the generation mix, in the structure of the network, and in the means and profiles of electricity consumption. The increased share of intermittent generation requires large amount of reserves and costly infrastructure expansions, while the electrification of energy consumption will significantly distort the aggregate electricity consumption profile. This is mainly due to EVs that consume as much as the rest of the loads in a typical household over a small charging period \cite{hashmi2018load}. The negative consequences of such distributed energy resources can be avoided by an adequate response from  active energy participants with flexible energy consumption and/or generation (prosumers). In order to achieve such a  response, incentives are provided to interested participants, who receive economic rewards in exchange of their flexibility services. Authors in \cite{hashmi2018effect} observe that with the growth of renewable share the need for  such responsive users is  going to increase. This represents an opportunity for energy participants to start providing services to the grid. Energy storage devices such as batteries are at the focal point of such applications, as these are gradually becoming profitable thanks to increasing flexibility opportunities and rewards and to continuous drop of their cost \cite{hashmi2019sizing}.

Authors in \cite{lam2015economics} present the economic analysis of storage in Southern California Homes. They highlight that earlier Net-Energy Metering or NEM policies allowed only excess renewable generation to be supplied back to the grid. However, new NEM policy allows distributed generation along with solar to participate in NEM making it more conducive for consumers to invest in energy storage.

For low voltage consumers, performing energy arbitrage is one of the prominent applications of storage devices. Optimizing storage for performing energy arbitrage is studied in numerous works, some of which are \cite{Lee2007}, \cite{nguyen2017maximizing}, \cite{hashmi2017optimal}.
In \cite{Lee2007}, the authors propose an algorithm mixture of dynamic
programming and particle swarm optimization to schedule a battery in presence of a wind turbine. The
time-of-use or ToU has 3 leaves and they manage to obtain a 5.6 \% saving in the energy bill
each month. 
Authors in \cite{nguyen2017maximizing} use linear programming for storage control under ToU pricing for PG\&E residential and consumers in San Francisco. 
Authors in \cite{hashmi2017optimal} present a convex optimization formulation for storage control for equal buying and selling rates of electricity. 
They also present a case-study for several energy markets in the USA and Europe and identify the storage financial potential in those energy markets. It is identified that only arbitrage is not financially viable due to high cost and limited life of the battery.

Co-optimizing energy storage for multiple revenue streams could drastically enhance the gains made by storage owners \cite{walawalkar2007economics}, \cite{shi2018using}, \cite{hashmi2018pfcpowertech}.
Pioneering work on co-optimization, \cite{walawalkar2007economics}, presents an application of batteries performing arbitrage and frequency regulation for a New York ISO case-study. They show that storage batteries might not be profitable for a single dedicated application due to its high cost.  Authors in \cite{shi2018using} claim that combining several applications could lead to greater gains compared to cumulative gains obtained by performing few of the tasks at a time. This is essential, as having one dedicated goal might maximize gains in that application but might lead to unexpected penalties  or undermine other applications. For instance storage performing only energy arbitrage could increase the peak demand charge paid by the consumer, in effect reducing the total gain made by the storage owner \cite{hashmi2018pfcpowertech}. A fine understanding of the billing mechanism is essential to avoid incurring such a penalty.

Profitability is an essential question from a consumer perspective, who needs to decide whether to invest in energy storage devices. 
The storage battery having a limited operational and aging life \cite{hashmi2018limiting}, economic analysis should consider the operational cycles in analyzing the gains made by using energy storage. 
Authors in \cite{Abdulla2018} observe that ignoring storage degradation could inflate the gains made by operating the storage battery. They also highlight the importance of forecasting parameters such as consumer load and electricity price for maximizing prosumer gains.
Authors in \cite{hashmi2019energy} present a case-study for Madeira island power network, where they explore  residential storage applications and their economic viability. They also provide recommendation about the best-suited contract among 46 different options for low voltage consumers based on short-term and long-term simulations of 1 day and 1 month respectively. 
In \cite{Carpenter2012}, the authors assess the profitability of storage in
Ontario under various pricing schemes, without the possibility to sell
energy back to the grid, showing that for several cases it cannot be achieved. 
Authors in \cite{Byrne2013} evaluate the feasibility of energy arbitrage
(without selling back to the grid) in Australia using time of use pricing. For
several batteries they found the probability of earnings depending on 
battery size.
Profitability of storage devices is governed by grid norms which represent the eagerness and needs of power networks to incentivize consumers for installing energy storage. The compensation mechanisms are also influenced by political will, as electricity is not just any commodity in a market for profit making but also a necessity in today's world.

%


%
%

\begin{figure*}
	\centering
	\includegraphics[width=5.2in]{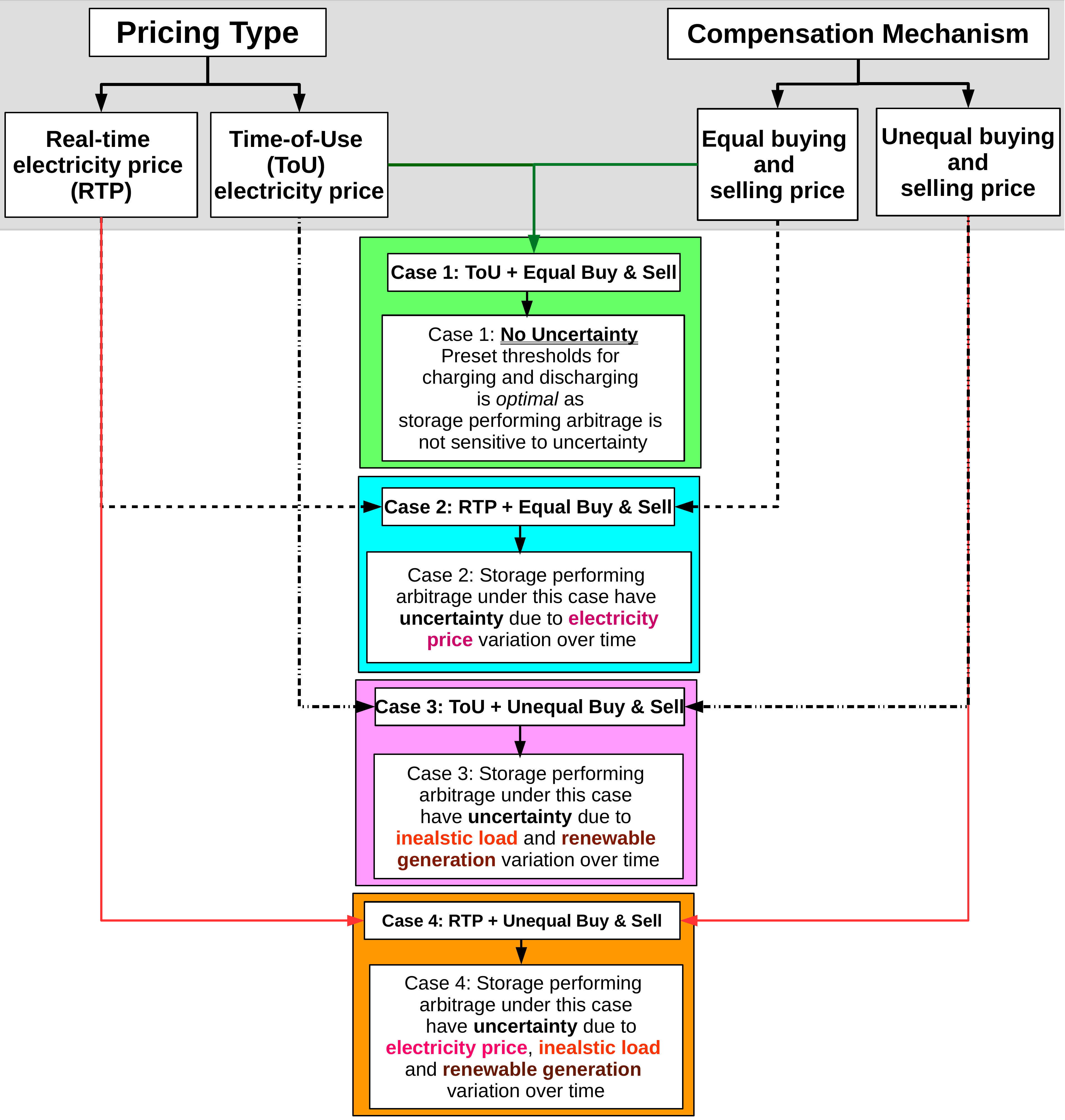}
	\caption{\small{Conditions and effect of parameter uncertain under storage performing arbitrage; Consider the consumer have an inelastic load, local renewable generation and storage. Under this setting storage operation is prone towards uncertainty depending on pricing type and excess generation compensation mechanism \cite{hashmi2019optimization}.}}\label{arbitrageanduncertainty}
\end{figure*}
In this work, we consider storage control and applications for LV consumers in Uruguay, where new contracts were proposed in early 2019, providing unique opportunities for consumers to include storage to modify the consumption profile seen by the utility meter. We present a co-optimization formulation for the energy storage device performing arbitrage and power factor correction (PFC), considering storage degradation. We find that some consumer contracts could be profitable enough for consumers to invest on storage. In \cite{hashmi2018pfc} authors identify that when co-optimizing storage for arbitrage and PFC, these objectives are largely decoupled as arbitrage is governed by storage charge level while  the reactive power capability is constrained by instantaneous active power and converter size. Furthermore, authors in \cite{hashmi2018pfc} also identify that lookahead for reactive power compensation is not essential, allowing us to consider myopic reactive compensation. For energy arbitrage, the control in the context of LV electricity consumer in Uruguay is simplified as

- \textit{No uncertainty in electricity price}: In Time-of-Use the instants and levels of price variations are known a priori thus no uncertainty in storage control due to electricity price. This is an advantage over real real-time electricity pricing schemes, where prices need to be forecasted.

- \textit{Net metering - Compensation of excess generation based on equal buying and selling price level}: Electricity buying price could vary over time, however, at every time instant the buying price and selling price remain the same in magnitude. For example, consider the buying price of electricity between 9 am and 10 am is 0.05\$/kWh; the consumer consumes say 20 kWh, making the cost of consumption equal to \$ 1. In the same time period had the prosumer instead of consuming generated 20 kWh of energy, he/she would have received compensation of \$ 1. Under such a compensation mechanism, the objective to minimize the cost of consumption of the user which includes inelastic load, renewable generation, and storage output is equivalent to maximizing the profit made by only energy storage performing arbitrage \cite{hashmi2017optimal}. In other words, optimal storage control decisions are independent of variations in inelastic load and renewable generation at consumer end, which was also observed in \cite{xu2017optimal}. Refer to Appendix~\ref{arbitragesec} for more details. Fig.~\ref{arbitrageanduncertainty} shows the effect of parameter uncertainty based on pricing mechanism and compensation of excess consumer generation. 
Thus, storage active power can be operated based on present thresholds which minimize the stress on the battery as the optimal solution. Note that because of the above-mentioned context, active power output of storage also does not require lookahead. The proposed consumer contracts make consumers completely immune to uncertainty in system variables, furthermore, no lookahead of such parameters are required. This should be very attractive for consumers as storage operation in a more dynamic market could affect consumer gains made by storage operation due to the real-time variation of system parameters causing a loss of opportunity. In \cite{yize2018stochastic} authors found that consumer gains can beup to 59\% lower for the stochastic case in comparison to the deterministic one. 
Furthermore, authors in \cite{krishnamurthy2018energy, thatte2013risk} propose storage control for performing arbitrage, taking into consideration parameter uncertainty in real time storage operation. These control mechanisms are computationally intensive, however, in our work we observe that energy storage operation becomes immune to parameter uncertainty due to pricing of electricity and excess generation compensation.

The paper is organized in sections. 
Section~\ref{sectionUrug2} provides a brief summary of the power system landscape in Uruguay. 
Section~\ref{sectionUrug3} presents the new electricity consumer contracts applicable for low voltage consumers in Uruguay.
Section~\ref{sectionUrug4} outlines the various applications energy storage can be used for the consumer contracts detailed in the previous section.
Section~\ref{sectionUrug5} describes the storage control algorithm applicable for different contracts. The proposed storage algorithm considers storage ramping and capacity constraint, charging and discharging efficiency losses.
Section~\ref{sectionUrug6} presents the numerical results using the storage control algorithm.
Section~\ref{sectionUrug7} concludes the paper.

\section{Energy Landscape in Uruguay}
\label{sectionUrug2}
Uruguay is a country with 3.5 million inhabitants located in the south of South
America, that has become a world leader in renewable energies. 
Historically, this small country has relied upon hydro power, mainly from two dams: Salto
Grande (co-owned with Argentina in the Uruguay river) and Rio Negro, thermal power
plants, and importing energy from neighboring countries: Argentina and Brazil.
As of today, Uruguay has completely changed their energy mix, consuming mostly
from renewable resources and becoming a purely exporting country.
In the year 2017, only 1\% of the energy was produced from thermal power stations,
while 65\% corresponded to hydro generation, and the remaining 34\% was
composed of a mix between biomass, wind and photovoltaic energy. With more than 600 wind turbines in 2018 wind power provided up to 49 \% of the energy consumed, which positions Uruguay as the second country in the world with the highest share of wind power in their electricity mix, behind Denmark.

The main actor in the energy sector is \href{https://portal.ute.com.uy/}{UTE}, a public company that acts as the only retailer, DSO and TSO, and it is also one of the main  producers of electricity. 
In a country with a surface of 176,215 $km^2$, the distribution and transmission lines span
84,245 km and 5561 km respectively as of 2017. As a matter of
comparison, France with 3.6 times the extension of Uruguay, has 18.8 times
more kilometers of transmission lines \cite{datarte}. 
Of the above mentioned mix, the residential sector consumes approximately 3500
GWh yearly, while big customers represent a load of 2000 GWh and medium
consumers account for 1500 kWh.
The peak power consumption registered in the years 2016 and 2017 was of 1964 MW
and 1916 MW respectively and occurred during the winter (mid July), while the yearly load factor was of 64.3 \%.

Uruguay's energy sector is still evolving towards a smarter landscape.
UTE has several demand response programs in place. Among them, they offer
time-of-use tariff to residential consumers, 60\% price rebates for big consumers who increase their demand during periods of renewable energy surplus, and price rebates on efficient household appliances such as class A water
heater. Regarding distribution of residential clients among tariffs, 47 \% of total clients subscribe to
a flat rate tariff while 53\% use a more complex contract such as time-of-use.

\section{Electricity Consumer Contracts}
\label{sectionUrug3}
Low Voltage (LV) electricity consumers can opt for new billing mechanisms introduced by UTE in January 2019 \cite{tariffute}, leading residential consumers to choose among three different consumer contracts.
Next we describe the three contracts denoted as C1, C2 and C3 in this document for easy referencing. They are structured with a variable fee that depends on the amount or the time of active energy consumption, a part that is proportional to the contracted peak power and a fixed charge that is applied to all contracts every month, irrespective of their variable energy or power consumption. 
Furthermore, consumers pay a reactive energy charge. The details of the contracts can be found Portal of Electricity operator in Uruguay\footnote{\url{https://tinyurl.com/y5ug28jh}}. The electricity prices are listed in Uruguayan peso\footnote{1 Uruguayan Peso equals 0.031 USA Dollar on 21st Feb. 2019.}.

This section is divided into five sub-sections. In Section~\ref{contractfixed} we list the fixed and variable active energy rates for the different contracts. In Section~\ref{contractpeak} the peak contracted power is given. Note that consumers need to select the peak power contracts beforehand, as it determines the cabling and metering requirements and therefore, do not have a flexibility in real-time to optimize. In turn, this will determine the fixed costs incurred by the consumer.
The billing of reactive power under the three contracts are presented in Section~\ref{reactivebil}.
Section~\ref{contracttotal} presents the total and variable cost of electricity consumption. In this paper, we use the variable component to minimize the cost of consumption of user using a battery.
Consumers with distributed generation such as rooftop solar generation can opt for net-metering in Uruguay. Section~\ref{contractnem} presents the net-metering policies for different consumer contracts.

\subsection{Fixed and Active Energy Cost}
\label{contractfixed}
Table~\ref{simplerate}, Table~\ref{tou2rate} and Table~\ref{tou3rate} lists the fixed, power and energy charges for contracts C1, C2 and C3. The fixed cost for contract $Ci$ is denoted as $C_{\text{fixed}}^{Ci}$ for $i \in \{1,2,3\}$.
We also present the active energy charge calculated under the different contracts for consumers.
\subsubsection{C1: Simple Residential Flat Rate}
Simple residential contracts are applicable for consumers with voltage level 230V and 400V and the contracted power is less than or equal to 40 kW.
\begin{table}
	\caption {\small{C1: Simple Residential Rate}}
	\label{simplerate}
	\vspace{-10pt}
	\begin{center}
		\begin{tabular}{| c | c| }
			\hline
			Category & Price\\ 
			\hline
			\hline
			\textbf{Charge for energy consumption}: & \\
			1 kWh to 100 kWh monthly &  5.160 peso/kWh \\
			101 kWh to 600 kWh monthly &  6.470 peso/kWh \\
			601 kWh onwards &  8.065 peso/kWh \\
			\hline
			\textbf{Charge for contracted power} &  61.6 peso/kW \\
			\hline
			\textbf{Fixed monthly charge} &  198.9 peso \\
			\hline
		\end{tabular}
		\hfill\
	\end{center}
\end{table}
The cost of active energy is given as:
\begin{equation}
C_{\text{active}}^{C1} = \lambda_{\text{fixed}} E_a,
\end{equation}
where $E_a$ denotes the active energy consumed and expressed in kWh and $\lambda_{\text{fixed}}$ denotes the flat rate electricity cost under C1.

\subsubsection{C2: Two level Time-of-Use (ToU) Rates}
Two level ToU residential contracts are applicable for consumers with voltage level 230V and 400V and the contracted power greater than 3.3 kW and less than or equal to 40 kW.
\begin{table}
	\caption {\small{C2: Two-level ToU Residential Rate}}
	\label{tou2rate}
	\vspace{-10pt}
	\begin{center}
		\begin{tabular}{| c | c| }
			\hline
			Category & Price\\ 
			\hline
			\hline
			\textbf{Charge for energy consumption}: & \\
			Peak hours: from 17:00  to 23:00 & 8.623 peso/kWh\\
			Off-peak hours: 00:00 to 17:00  & 3.453 peso/kWh\\
			and 23:00 to 24:00 hrs & \\
			\hline
			\textbf{Charge for contracted power} &  61.6 peso/kW \\
			\hline
			\textbf{Fixed monthly charge} &  359.4 peso \\
			\hline
		\end{tabular}
		\hfill\
	\end{center}
\end{table}
The cost of active energy is given as:
\begin{equation}
C_{\text{active}}^{C2} = \lambda_{\text{peak}} E_a^{\text{peak}} + \lambda_{\text{off-peak}} E_a^{\text{off-peak}},
\end{equation}
where $E_a^{\text{peak}}$ denotes the active energy consumed during peak period over the month,
$E_a^{\text{off-peak}}$ denotes the active energy consumed during off-peak period over the month,
and expressed in kWh and $\lambda_{\text{peak}}$ denotes peak electricity cost, $\lambda_{\text{off-peak}}$ denotes off-peak electricity cost under C2 contract.

\subsubsection{C3: Three-level ToU Rates}
Three level ToU residential contracts are applicable for consumers with voltage level 230V and 400V and the contracted power greater than 3.7 kW and less than or equal to 40 kW.
\begin{table}
	\caption {\small{C3: Three-level ToU Residential Rate}}
	\label{tou3rate}
	\vspace{-10pt}
	\begin{center}
		\begin{tabular}{| c | c| }
			\hline
			Category & Price\\ 
			\hline
			\hline
			\textbf{Charge for energy consumption}: & \\
			Peak hours: from 17:00 to 23:00 hrs & 8.623 peso/kWh\\
			Mid-peak hours: 07:00 to 17:00 & 4.676 peso/kWh\\
			and 23:00 to 24:00 hrs & \\
			Off-peak hours: 00:00 to 7:00 hrs & 1.803 peso/kWh\\
			\hline
			\textbf{Charge for contracted power} &  61.6 peso/kW \\
			\hline
			\textbf{Fixed monthly charge} &  359.4 peso \\
			\hline
		\end{tabular}
		\hfill\
	\end{center}
\end{table}
The cost of active energy is given as:
\begin{equation}
C_{\text{active}}^{C3} = \lambda_{\text{peak}} E_a^{\text{peak}} + \lambda_{\text{mid-peak}} E_a^{\text{mid-peak}} + \lambda_{\text{off-peak}} E_a^{\text{off-peak}},
\end{equation}
where 
$E_a^{\text{mid-peak}}$ denotes the active energy consumed during mid-peak period over the month
and expressed in kWh and $\lambda_{\text{mid-peak}}$ denotes mid-peak electricity cost under C3 contract.

\subsection{Peak Power Contract for LV Consumers}
\label{contractpeak}
For low voltage consumers in Uruguay should specify the peak power contracted. This is essential as the utility provides the connection and safety features based on the contracted power.

Single phase consumers in LV network can select the contracted power from following levels presented in Table~\ref{ppcurg1}.  
Three phase consumers in LV network can select the contracted power from following levels presented in Table~\ref{ppcurg2}.
Based on Table~\ref{ppcurg1} and Table~\ref{ppcurg2} consumers select their power contract denoted as $P_{\text{contracted}}$. The charge for power contracted for contract $i \in \{1,2,3\}$ is denoted as $C_{\text{power}}^{Ci}$.	
The cost of power contracted is given as:
\begin{equation}
C_{\text{power}}^{Ci} = \lambda_{\text{power}} P_{\text{contracted}}.
\end{equation}		
The value of $\lambda_{\text{power}}$ is listed in Table~\ref{simplerate}, Table~\ref{tou2rate} and Table~\ref{tou3rate} for contracts C1, C2 and C3 respectively.
Note the peak power charge is same for C1/C2/C3.
\begin{table}
	\caption {\small{Power Contracted: 1-phase LV consumers}}
	\label{ppcurg1}
	\vspace{-10pt}
	\begin{center}
		\begin{tabular}{| c |  }
			\hline
			Power levels\\ 
			\hline
			\hline
			3.7 kW, 	4.6 kW,	7.4 kW, 9.2 kW \\
			\hline
		\end{tabular}
		\hfill\
	\end{center}
\end{table}
\begin{table}
	\caption {\small{Power Contracted: 3-phase LV consumers}}
	\label{ppcurg2}
	\vspace{-10pt}
	\begin{center}
		\begin{tabular}{| c |  }
			\hline
			Power levels (kW)\\ 
			\hline
			\hline
			12, 20, 25, 30, 35, \\
			40 kW, 41 to 50 kW\\
			\hline
		\end{tabular}
		\hfill\
	\end{center}
\end{table}


\subsection{Billing of Reactive Energy}
\label{reactivebil}
Traditionally, LV consumers were not obliged to regulate reactive power. There where many well thought reasons why it made sense, we list a few below:\\
$\bullet$  Majority of loads used by low voltage consumers consisted of resistive loads, thus, the inherent power factor seen by the grid used to be close to unity.\\
$\bullet$  Utility neither had the infrastructure nor the motivation for making it obligatory for small LV consumers to comply reactive power norms normally well-defined for commercial establishments.

However, the reasons why utilities across the world do not meter reactive power in LV networks is rapidly changing due to some of the following reasons:\\
$\bullet$  Evolution of many new loads have significantly increased the reactive power of the LV consumers. In absence of regulation utilities would have to face degradation of efficiency of LV distribution network \cite{pfc2007} and additional stress on distribution transformer. \\
$\bullet$  Most countries have been promoting distributed generation (DG) such as rooftop solar PV installations. These DGs operate at close to unity power factor, implying that while an important part of the active power is locally met by the DG, all the reactive power is provided/absorbed by the grid \cite{hashmi2018pfc}.

Due the above mentioned transformation, the utilities are designing penalties for consumers with a low power factor. Uruguay is in the fore front globally in implementing reactive power penalties and incentives for small LV consumers to control their power factor.

Next we describe the mechanisms of charging for degraded power factor for each of the three consumer contracts.

\subsubsection{Reactive power cost under C1}

Consumers billed in accordance to C1, or simple residential contract, need to pay a penalty if the power factor calculated for the month deteriorates below 0.92.
The power factor is calculated as a function of aggregate reactive power ($E_r$) and active power ($E_a$) as follows
\begin{equation}
\text{pf}_{\text{month}} = \cos\Big(\arctan\left(\frac{E_r}{E_a}\right)\Big).
\end{equation}
The cost or reactive power for contract C1 is given as
\begin{equation}
C_{\text{reactive}}^{C1} =  K_{\text{fac}} \times E_r.
\end{equation}
Where $K_{\text{fac}}$ is the coefficient of surcharge for reactive consumption and is governed by the following conditions:
\begin{equation}
K_{\text{fac}} =
\begin{cases}
0, & \text{if }  \frac{E_r}{E_a} \leq 0.426, \\
0.4 \left(\frac{E_r}{E_a} - 0.426 \right),& \text{if }   \frac{E_r}{E_a} \in (0.426,~ 0.7] ,\\
0.4 \left(\frac{E_r}{E_a} - 0.426 \right) + 0.6 \left(\frac{E_r}{E_a} - 0.7 \right) ,& \text{if }  \frac{E_r}{E_a} > 0.7. \\
\end{cases}
\label{eq:c1}
\end{equation}
\begin{figure}
	\centering
	\includegraphics[width=3.1in]{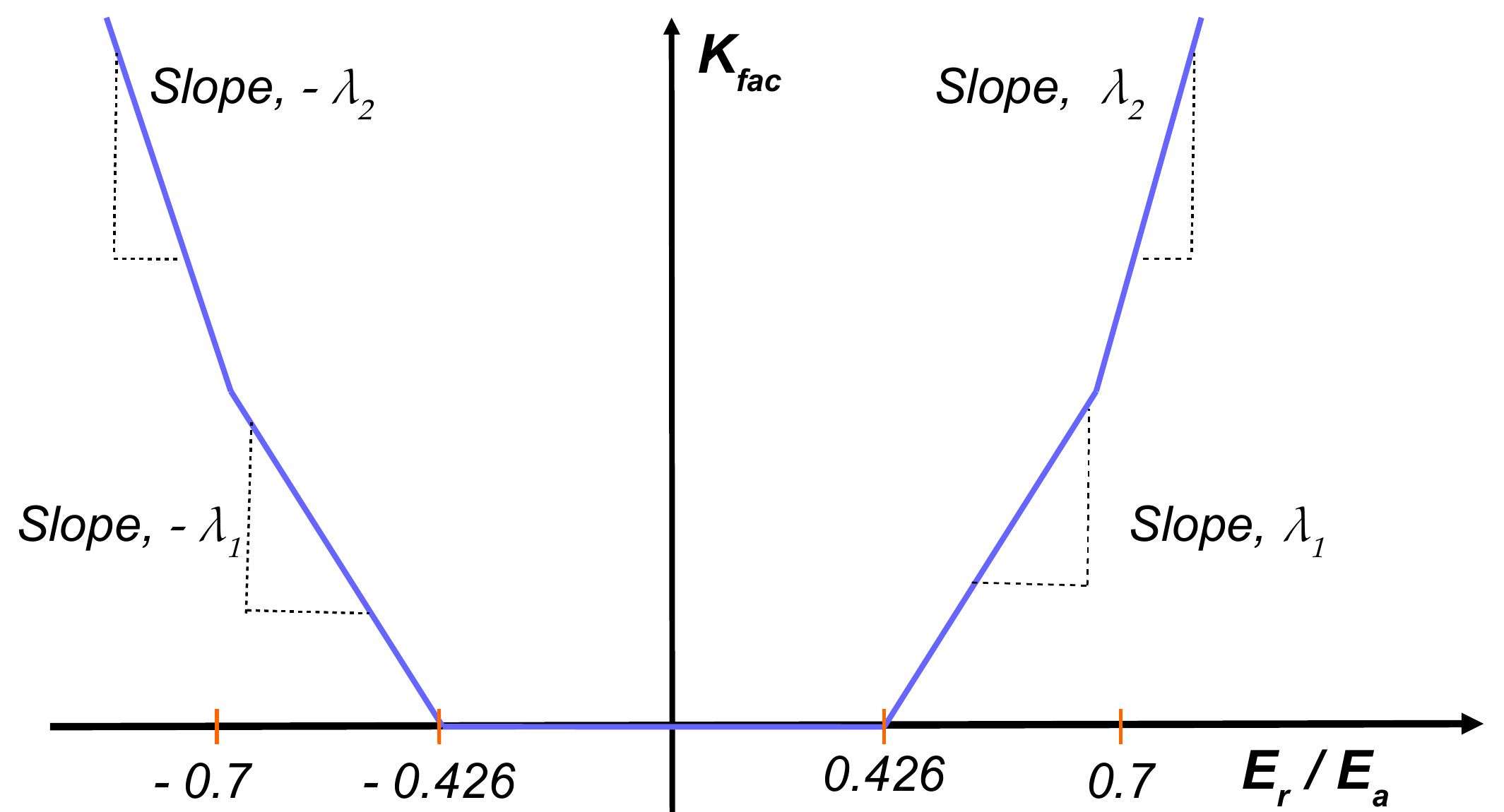} \vspace{-7pt}
	\caption{Graphical representation of $K_{\text{fac}}$ for C1 with respect to the PF; where $\lambda_1 = 0.4$ and $\lambda_2 = 1$.}\label{sysc1}
\end{figure}
\subsubsection{Reactive power cost under C2}
Consumers billed in accordance to C2, or two-level ToU residential contract, need to pay a penalty if the power factor calculated for the month deteriorates below 0.92. Consumers under C2 are also provided incentives for cases where the $\text{pf}_{\text{month}}$ exceeds 0.92.
The coefficient of surcharge or consumption bonus for reactive consumption is governed by following conditions:
\begin{equation}
K_{\text{fac}} =
\begin{cases}
\frac{B}{100} \left(\frac{E_r}{E_a} - 0.426 \right), & \text{if }  \frac{E_r}{E_a} \leq 0.7, \\
\frac{B}{100} \left(\frac{E_r}{E_a} - 0.426 \right) + \frac{100-B}{100} \left(\frac{E_r}{E_a} - 0.7 \right) ,& \text{if }  \frac{E_r}{E_a} > 0.7, \\
\end{cases}
\label{eq:c2}
\end{equation}
where the value of B is 36 or 34 depending on the contract.
For consumers under C2, the coefficient $K_{\text{fac}}$ acts as penalty for case where 
$\frac{E_r}{E_a} > 0.426$. However, if $\frac{E_r}{E_a} \leq 0.426$ then it provides incentives for the consumer as the cost of reactive power would be negative. 
Coefficient $K_{\text{fac}}$ is applied to the total active energy consumed during the peak period over the whole month. The total active energy is denoted as
\begin{equation}
E_a = E_a^{\text{off-peak}} + E_a^{\text{peak}},
\end{equation}
The cost or reactive power for contract C2 is given as
\begin{equation}
C_{\text{reactive}}^{C2} =  K_{\text{fac}} \times E_a^{\text{peak}}.
\end{equation}

\begin{figure}
	\centering
	\includegraphics[width=3.2in]{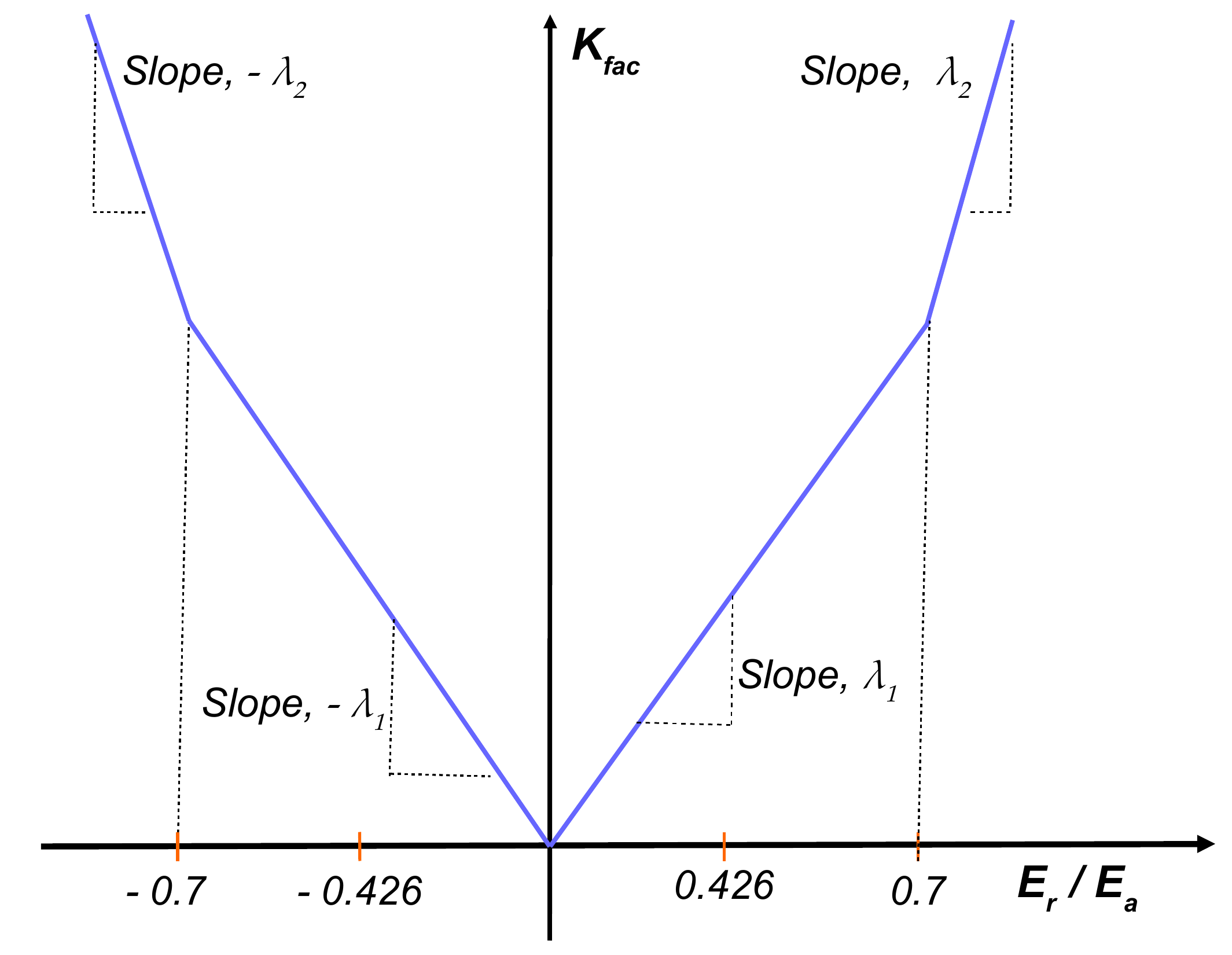} \vspace{-7pt}
	\caption{Graphical representation of $K_{\text{fac}}$ for C2 with respect to the PF; where $\lambda_1 = \frac{B}{100}$ and $\lambda_2 = 1$.}\label{chUsys}
\end{figure}

\subsubsection{Reactive power cost under C3}

The power factor applied for C3 consumers is given as
\begin{equation}
\text{pf}_{\text{month}}^+ = \cos\Big(\arctan\left(\frac{E_{rQ1}}{E_{a+}}\right)\Big),
\end{equation}
where $E_{rQ1}$ is the absolute value of reactive energy in Quadrant 1 over a month expressed in kVAR and $E_{a+}$ denotes the absolute value of active energy in the month expressed in units kWh.
Consumers billed in accordance to C3, or three level ToU residential contract, need to pay a penalty if the power factor, calculated with aggregate reactive power ($E_{rQ1}$) and active power ($E_{a+}$), calculated for the month deteriorates below 0.92.
Similar to C2, 
Consumers under C3 are also provided incentives for cases where the $\text{pf}_{\text{month}}^+ $ exceeds 0.92.

The coefficient of surcharge or consumption bonus for reactive consumption is governed by following conditions:

\begin{equation}
K_{\text{fac}} =
\begin{cases}
\frac{A}{100} \left(\frac{E_{rQ1}}{E_{a+}} - 0.426 \right), & \text{if }  \frac{E_{rQ1}}{E_{a+}} \leq 0.7, \\
\frac{A}{100} \left(\frac{E_{rQ1}}{E_{a+}} - 0.426\right) + \frac{100-B}{100} (\frac{E_{rQ1}}{E_{a+}} - 0.7) ,& \text{if }  \frac{E_{rQ1}}{E_{a+}} > 0.7, \\
\end{cases}
\label{eq:c3}
\end{equation}
where the value of B depends on supply voltage listed in Table~\ref{valueA}.
\begin{table}
	\caption {\small{Value of B under C3}}
	\label{valueA}
	\vspace{-10pt}
	\begin{center}
		\begin{tabular}{| c | c| }
			\hline
			Voltage level & Value of B\\ 
			\hline
			\hline
			230 to 400 V & 23\\
			6.4 - 12 - 22 kV & 18 \\
			31.5 kV & 12\\
			\hline
		\end{tabular}
		\hfill\
	\end{center}
\end{table}
Coefficient $K_{\text{fac}}$ is applied to the total active energy consumed during the peak period over the whole month. The total active energy is denoted as
\begin{equation}
E_{a+} =  E_{a+}^{\text{off peak}} + E_{a+}^{\text{mid peak}}+ E_{a+}^{\text{peak}},
\end{equation}
where $E_{a+}^{\text{off peak}}$ denotes the total energy consumed during off-peak hours over the month, $E_{a+}^{\text{mid peak}}$ denotes the total energy consumed during mid-peak hours over the month and $E_{a+}^{\text{peak}}$ denotes the total energy consumed during peak hours over the month.
The cost or reactive power for contract C3 is given as
\begin{equation}
C_{\text{reactive}}^{C3} =  K_{\text{fac}} \times E_{a+}.
\end{equation}

\begin{figure}
	\centering
	\includegraphics[width=3.2in]{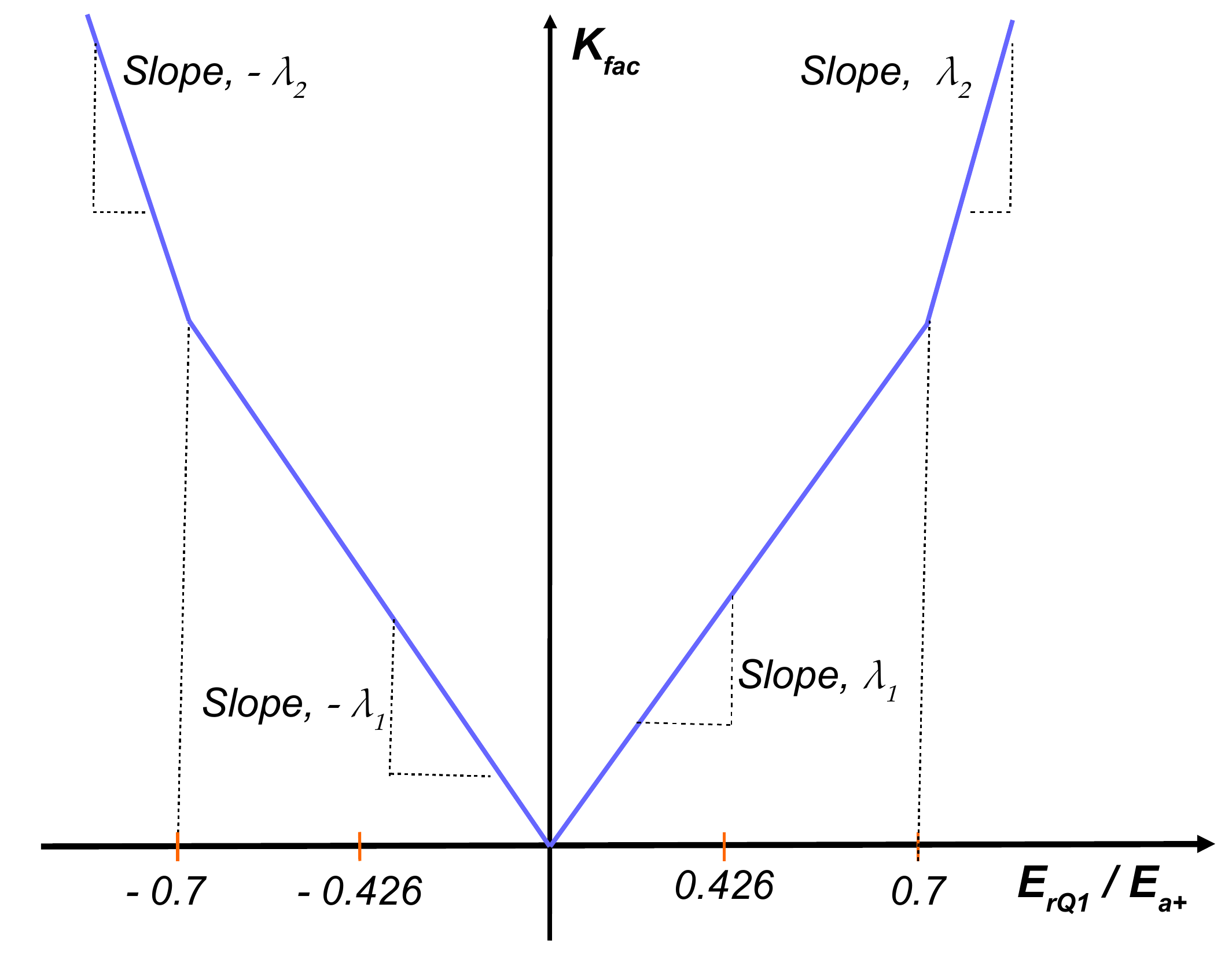} \vspace{-7pt}
	\caption{Graphical representation of $K_{\text{fac}}$ for C3 with respect to the PF; where $\lambda_1 = \frac{A}{100}$ and $\lambda_2 = 1$.}\label{sysc3}
\end{figure}

\subsection{Cost of Consumption}
\label{contracttotal}
The cost of electricity consists of four components:

(1.) \textit{Fixed electricity cost}: depending on the contract type, consumers are charged a fixed cost which is independent of the consumed electricity,

(2.) \textit{Power contracted cost}: the consumer specifies the peak power level before the utility provides a connection. From the utility perspectives this contract determines the protection settings, fuse settings, cable ratings, meter type etc.

(3.) \textit{Active energy cost}: The cost of consuming active energy is decided by the contract type.

(4.) \textit{Reactive energy cost}: This component of electricity cost is decided by the power factor, voltage level, total active energy consumed and contract type.

Therefore, the total cost of consumption under contract $Ci$ is given as
\begin{equation}
C_{\text{Total}}^{Ci} = C_{\text{fixed}}^{Ci} + C_{\text{power}}^{Ci} + C_{\text{active}}^{Ci} + C_{\text{reactive}}^{Ci}.
\end{equation}

The variable component of the cost would consists of the cost of active and reactive power given as
\begin{equation}
C_{\text{variable}}^{Ci} = C_{\text{active}}^{Ci} + C_{\text{reactive}}^{Ci}.
\end{equation}

The variable component of electricity price could only be reduced if the electricity consumer optimizes their consumption locally using load flexibility and/or energy storage. The cost component $C_{\text{fixed}}^{Ci} + C_{\text{power}}^{Ci}$ have no degree of freedom and therefore, cannot be reduced.
\begin{table}
	\caption {\small{Variable cost component based on contract type}}
	\label{variablecost}
	\vspace{-10pt}
	\begin{center}
		\begin{tabular}{| c | c| }
			\hline
			Contract & $C_{\text{variable}}^{Ci}$\vspace{1mm}\\ 
			\hline
			\hline
			C1 & $\lambda_{\text{fixed}} E_a  +   K_{\text{fac}} \times E_r  $ \vspace{2mm} \\ 
			C2 & $\lambda_{\text{peak}} E_a^{\text{peak}} + \lambda_{\text{off-peak}} E_a^{\text{off-peak}}   +   $\\
			& $K_{\text{fac}} \times E_a^{\text{peak}}$ 	 \vspace{2mm}\\
			C3 & $\lambda_{\text{peak}} E_a^{\text{peak}} + \lambda_{\text{mid-peak}} E_a^{\text{mid-peak}} + $\\
			& $\lambda_{\text{off-peak}} E_a^{\text{off-peak}} +     K_{\text{fac}} \times E_{a+}^{\text{peak}}$\\
			\hline
		\end{tabular}
		\hfill\
	\end{center}
\end{table}
\subsection{Net-Metering in Uruguay}
\label{contractnem}
\begin{figure}
	\centering
	\includegraphics[width=3.3in]{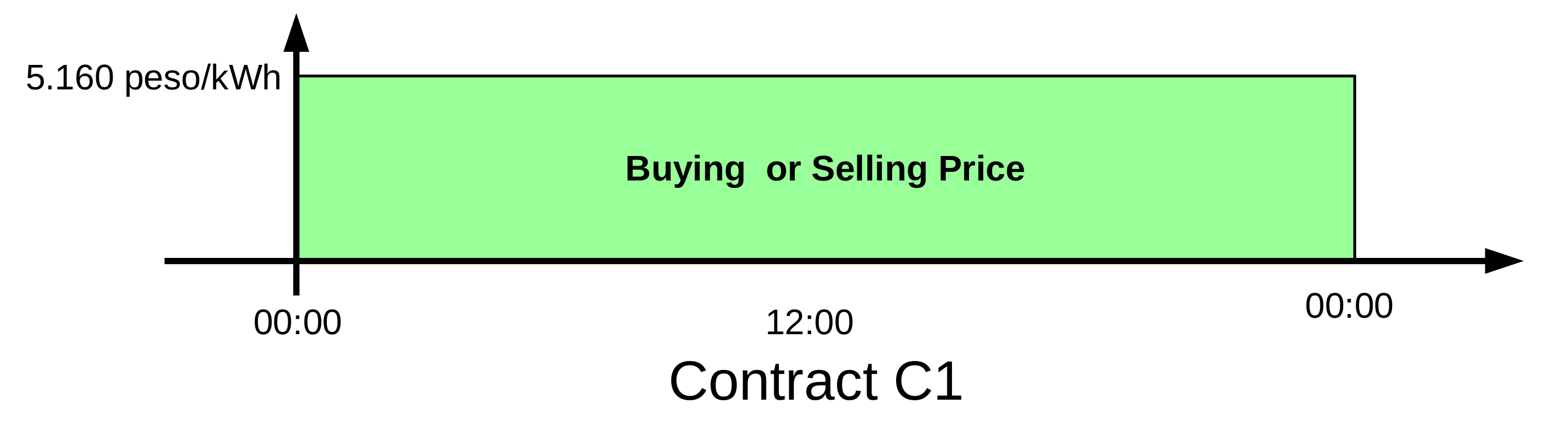}
	\vspace{-5pt}
	\caption{Buying and selling price of electricity over a day under contract C1 with NEM}\label{c1nem}
\end{figure}

\begin{figure}
	\centering
	\includegraphics[width=3.3in]{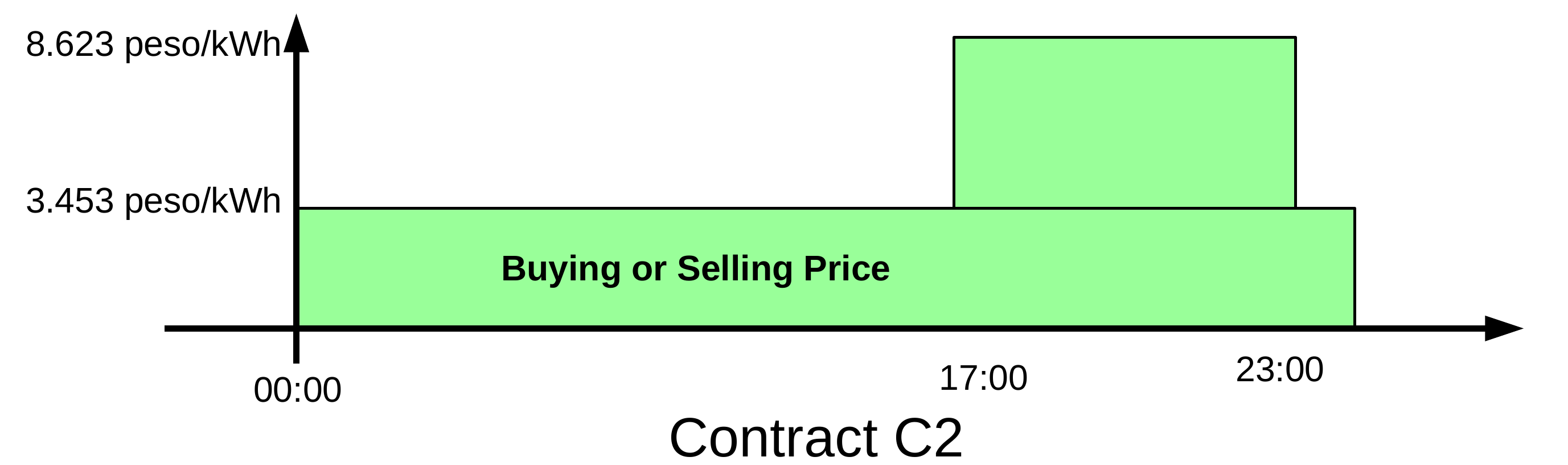}
	\vspace{-5pt}
	\caption{Buying and selling price of electricity over a day under contract C2 with NEM}\label{c2nem}
\end{figure}

\begin{figure}
	\centering
	\includegraphics[width=3.3in]{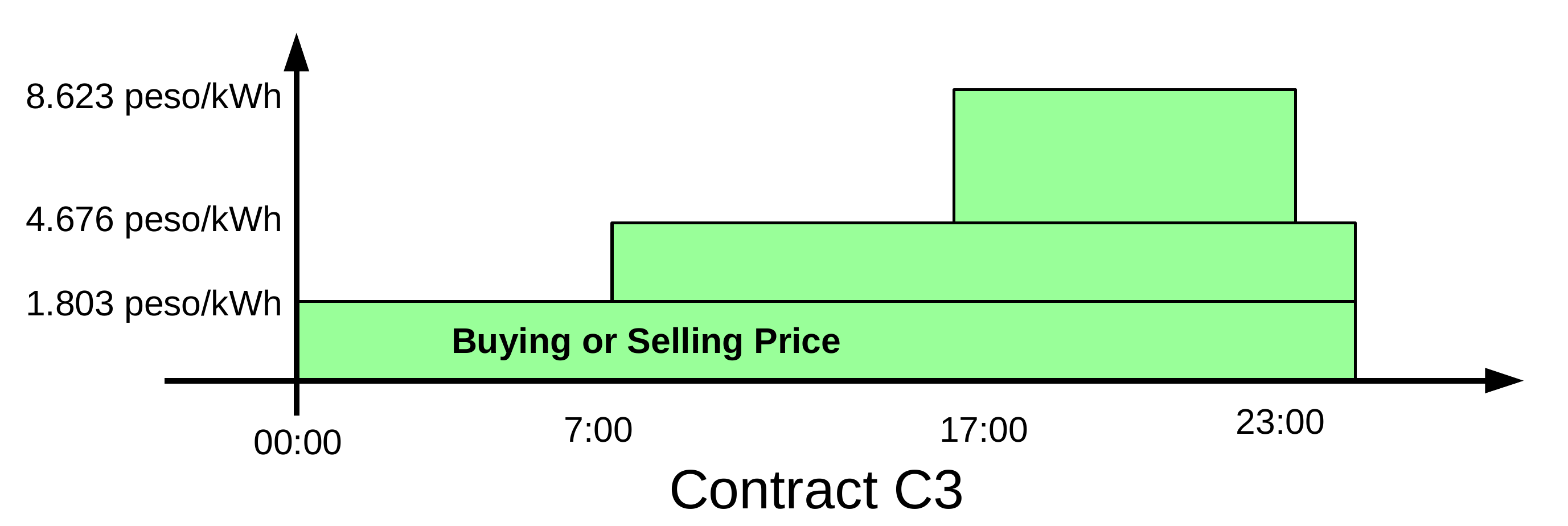}
	\vspace{-5pt}
	\caption{Buying and selling price of electricity over a day under contract C3 with NEM}\label{c3nem}
\end{figure}

Net-energy metering (NEM) for small wind power, solar,
biomass and mini-hydro systems is allowed
since 2010 by Decree 173/010\footnote{\url{http://tinyurl.com/y2ccr2o2}}
on
micro-generation. The government-owned
national electric company, UTE\footnote{\url{https://portal.ute.com.uy/}}, is mandated to buy at
retail price all the excess electricity produced
by consumers for a period of ten years \cite{uruguaynem}.
Generated electricity must be low-voltage and 
the maximum power of installations is the
lower between 6 kW and the peak power contracted
by consumer, although higher power is
possible with additional authorization before installation\footnote{\url{https://tinyurl.com/y2gydsu7}}.
The buying and selling electricity prices for LV consumers in Uruguay under contracts C1, C2 and C3 are depicted in Fig.~\ref{c1nem}, Fig.~\ref{c2nem} and Fig.~\ref{c3nem} which shows that buy and sell price have the same level. {Note that consumer needs to opt for NEM separately by notifying the utility which may require the installation of different hardware.}


In this paper we use net-metering for contracts C2 and C3. The rationale behind not opting for net-metering for contract C1 is described in Section~\ref{contractc1nem}.

\section{Storage for LV Prosumers in Uruguay}
\label{sectionUrug4}
\begin{figure}
	\centering
	\includegraphics[width=3.2in]{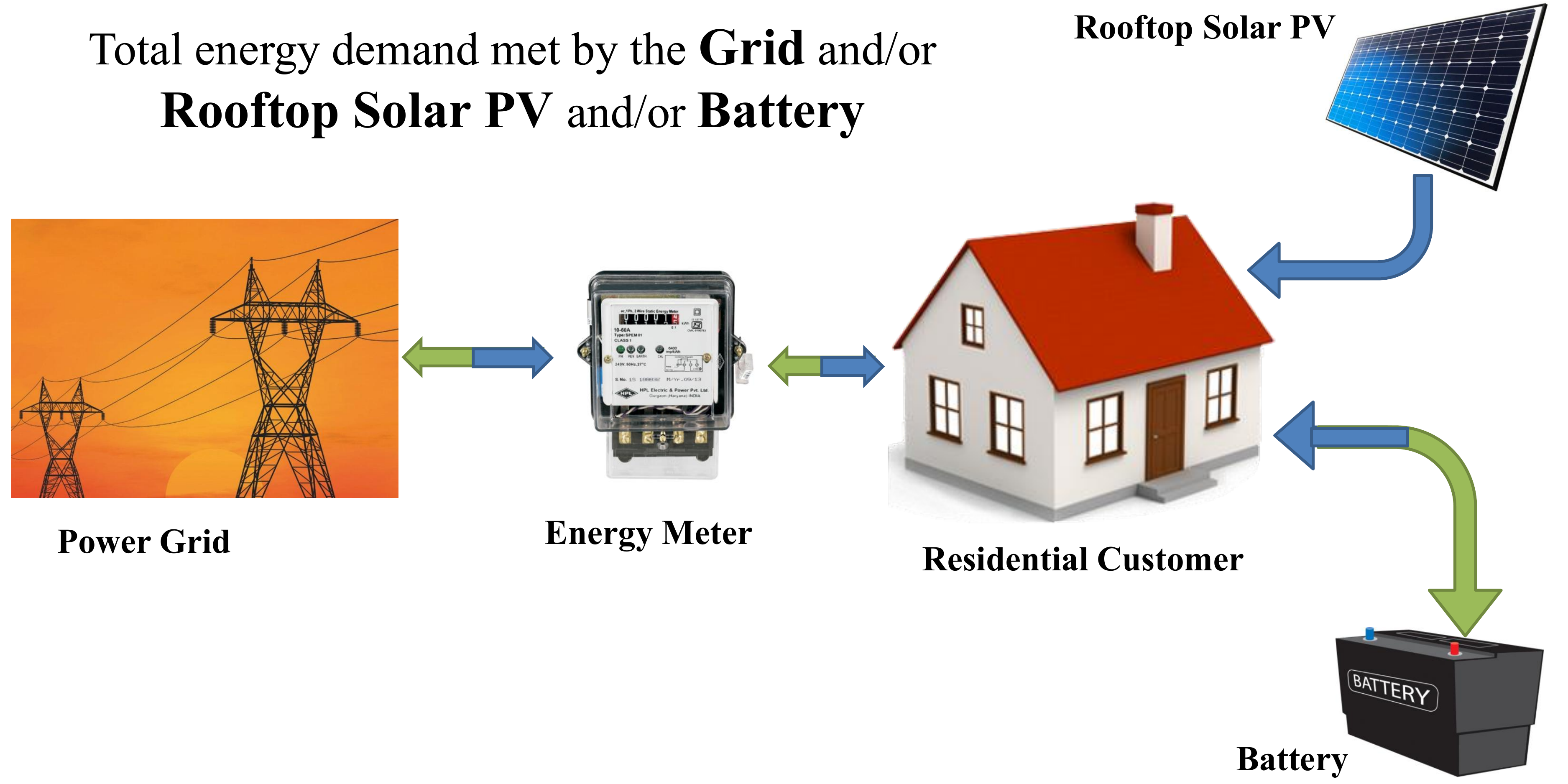}
	\vspace{-10pt}
	\caption{System Considered}\label{case0}
\end{figure}

The system considered in this work consists of an electricity consumer with inelastic demand, renewable generation (rooftop solar) and energy storage. The battery will provide flexibility to deviate consumption in order to make gains by performing arbitrage and provide reactive energy compensation. 
The system is shown in Fig.~\ref{case0}.
The energy storage interfaced via a converter provides flexibility to the consumer to modify the active and reactive power seen by the grid. Energy storage with converter can act as source and/or sink of active and reactive power. 
Opportunities for prosumers with storage in Uruguay are:

 \textbf{Arbitrage}: the ToU cost structure makes the selection of arbitrage decisions fairly simple for storage owners. Furthermore, since under Net-metering policies in Uruguay, the buy and sell price is equal for each time instant, shown in Fig.~\ref{c1nem}, Fig.~\ref{c2nem} and Fig.~\ref{c3nem}, this implies that the control energy storage becomes independent of the inelastic load and solar generation.
	
	From the structure of ToU prices, clearly the storage needs to charge during the off-peak period and discharge during the peak period. Storage remains idle during mid-peak periods except for very slow ramping batteries which could not be completely charged in off-peak duration or discharged completely during peak period duration.

 \textbf{Avoiding power factor penalty}: previously we described the billing mechanism for reactive power for LV consumers in Section~\ref{reactivebil}. The thresholds indicate that as power factor (PF) seen by the grid deteriorates the cost of consumption increases in proportion to peak energy consumed; this penalty for low PF could be avoided by maintaining the PF locally. Energy storage interfaced via a converter could be used for PF improvement locally.

 \textbf{Maintaining a high PF} so as the consumer receives additional rebate on electricity consumption cost. Note that the consumers maintaining a unity PF could reduce the active power cost under C2 by almost 13.6\% of the cost energy consumed under the peak period of ToU. For contract C3, the  cost of consumption could be reduced by utmost 9.8\% of the cost of energy consumed during peak period. However, for C1, there is no additional incentive for consumers improving the PF above 0.92. Users opting for C1 could make profit by reactive power compensation if the PF without any such correction is lower than 0.92.

\subsection{Notation and Battery Model}
The inelastic load for time $i$ is denoted as $d_i$.
The distributed generation for time $i$ is denoted as $r_i$.
The total active load seen by the energy meter without storage is denoted as
$
P^i = d_i - r_i.
$
The reactive power at time $i$ without storage is denoted as $Q^i$.
We denote the battery using ramping and capacity constraint, considering charging and discharging efficiency. The ramping constraint is denoted as
\begin{equation}
\delta_i \in [\delta_{\min}, \delta_{\max}],
\end{equation}
where $\delta_{\min}$ denotes minimum ramping rate or the maximum discharge rate and $\delta_{\max}$ denotes the maximum charge rate in units of watts (i.e. power).

The instantaneous battery capacity is denoted as $b_i$ for time instant $i$. The battery capacity should satisfy the capacity constraint given as
\begin{equation}
b_i \in [b_{\min}, b_{\max}],
\end{equation}
where $b_{\min}$ and $b_{\max}$ denotes the minimum and maximum battery capacity. The instantaneous battery evolution with time depends the ramp rate of the battery and the battery capacity in the previous time instant and is given as
\begin{equation}
b_i =  b_{i-1} + x_i ,
\end{equation}
where $x_i = \delta_i h$ denotes the change in charge level of the battery. $h$ denotes the sampling time.
The state-of-charge (SoC) of the battery is defined as 
\begin{equation}
\text{SoC}_{i} = \frac{b_i}{b_{\text{rated}}},
\end{equation}
where $b_{\text{rated}}$ is the rated battery capacity. For LiIon battery health consideration, it should not be over-charged or over-discharged beyond a certain level \cite{buvsic2017distributed}, we define $\text{SoC}_{\min}=b_{\min}/b_{\text{rated}}$ and $\text{SoC}_{\max}=b_{\max}/b_{\text{rated}}$.

We define two more variables. The time required for the energy storage to completely discharge from $b_{\max}$ to $b_{\min}$ at the maximum discharge rate is given as
\begin{equation}
T_{dis} = \frac{b_{\max} - b_{\min}}{|\delta_{\min}|}.
\end{equation}

The time required for the energy storage to completely charge from $b_{\min}$ to $b_{\max}$ at the maximum charge rate is given as
\begin{equation}
T_{ch} = \frac{b_{\max} - b_{\min}}{\delta_{\max}}.
\end{equation}

The active power output of the battery is denoted as 
\begin{equation}
P_B^i = \frac{[x_i]^+}{h~\eta_{ch}} - \frac{[x_i]^-\eta_{dis}}{h},
\end{equation}
where $\eta_{ch}$ and $\eta_{dis}$ denotes charging and discharging efficiency of the battery and lies in the range $(0,1]$.
The active power ramp rate constraint is given as
\begin{align}
\footnotesize{
	P_B^i \in [P_B^{\min}, P_B^{\max}]\quad \text{~with~}P_B^{\min}\text{=}\delta_{\min}\eta_{dis}, P_B^{\max}\text{=}\frac{\delta_{\max}}{\eta_{ch}}}\label{constraintramp}
\end{align}

	Though the battery charge level is not affected by the reactive power output $Q_B^i$ of the connected inverter, the amount of active power supplied or consumed is dependent upon it due to the line current limitations \cite{kisacikoglu2011reactive}. 
	The converter rating is given by maximum apparent power supplied/consumed denoted as $S_B^{\max}$.
	The instantaneous apparent power of battery $S_B^i$ should satisfy
	\begin{equation}
	(S_B^{\max})^2 \geq {(P_B^i)^2 + (Q_B^i)^2}, 
	\label{constraintreactive}
	\end{equation}

%
The PF seen by the grid is the ratio of real power supplied or extracted by the grid over the apparent power seen by the grid. In the absence of storage it is given by 
\begin{equation}
\text{pf}_{\text{bc}}^i = \frac{P^i}{\sqrt{{P^i}^2 + {Q^i}^2}}.
\label{uncorrect}
\end{equation}
Observe that $\text{pf}_{\text{bc}}$ degrades as $r_i$ and $Q^i$ increases in magnitude.
In the presence of storage, PF takes the form 
\begin{equation}
pf_{\text{c}}^i = \frac{P^i + P_B^i}{\sqrt{(P^i + P_B^i)^2 + (Q^i+ Q_B^i)^2}}.
\label{correct2}
\end{equation}
where total active power is denoted as $P_T^i=P^i + P_B^i$ and total reactive power is given as $Q_T^i=Q^i + Q_B^i$.

\textit{Stress on battery}: Authors in \cite{mathieu2016controlling, fortenbacher2017modeling} identify that the internal storage losses depends square of the current supplied by the battery. Thus the storage operational profile which reduces the sum of square of currents over the time horizon would minimize the losses and thus we consider reduce the stress on the battery.


\subsection{Active Power Management}
Energy storage active power could serve two purposes: (1.) increase self-consumption of locally generated energy under the absence of NEM and (2.) perform arbitrage under ToU electricity pricing. 
\subsubsection{Under Contract C1}
\label{contractc1nem}
Energy price is fixed for the day and, therefore, no arbitrage is possible. Under fixed electricity price storage could maximize self-consumption locally in absence of NEM. However, self-consumption could only be increased for cases where the instantaneous generation is more than the instantaneous load \cite{hashmi2019energy}.

		\begin{table*}
			\caption {\small{Storage Operation under C1 without NEM}}
			\label{storagec1}
			\vspace{-10pt}
			\begin{center}
				\begin{tabular}{| c | c| c | c|}
					\hline
					Min Load & Max Load  & Installed & Storage\\ 
					without solar& without solar & Solar kWp& Role\\ 
					\hline
					\hline
					$L_{\min}$ & $L_{\max}$ & below $L_{\min}$ & No storage active energy management required\\
					$L_{\min}$ & $L_{\max}$ & between $L_{\min}$ and $L_{\max}$ & Storage might increases self-consumption\\
					$L_{\min}$ & $L_{\max}$ & above $L_{\max}$ & Storage increases self-consumption\\
					\hline
				\end{tabular}
				\hfill\
			\end{center}
		\end{table*}
		
Table~\ref{storagec1} lists the effect on self-consumption with the size of installed solar in absence of net-metering, which implies consumers have no incentive to supply power back to the grid and therefore, would attempt to increase the self-consumption as much as possible \cite{hashmi2019energy}. In Table~\ref{storagec1} $L_{\min}$, $L_{\max}$ denotes the minimum and maximum load without solar, $0 \leq L_{\min} \leq L_{\max}$. For case where installed solar is comparable or larger than maximum load without solar generation, storage substantially contributes in increasing self-consumption.

\begin{theorem}
	\label{profitabilityarb}
	For arbitrage to be profitable following condition should be valid:
	\begin{gather*}
	p_s^j \eta_{\text{dis}} > \frac{p_b^i}{\eta_{\text{ch}}},
	\end{gather*}
	where $p_s^j$ denotes the selling price at time instant $j$ and $p_b^i$ denotes the buying price at time instant $i$, such that $j>i$ and $j, i \in \{1,2,,...,N\}$. Here $N$ denotes the number of time instants in the horizon.
\end{theorem}
The proof of Theorem~\ref{profitabilityarb} is provided in Appendix~\ref{theorem41}.

Performing arbitrage with contract C1 with NEM will not be profitable as Theorem~\ref{profitabilityarb} is not valid, since $p_s^j = p_b^i~ \forall j,i$ and $\eta_{\text{ch}}, \eta_{\text{dis}} <1$.
In this work we explore the usage of energy storage for different contracts. 
Optimizing the energy storage is essential pertaining to its high cost. 
In context of C1 storage can only be used either for backing up excess generation and/or for peak demand shaving.
Backing up energy will require no look-ahead and greedy behavior leads to optimality \cite{hashmi2019energy}. The optimal solution in such a case is governed by the sign of $P^i$ and is given as\\
$\bullet$ If $P^i \geq 0$ then battery should discharge such that 
	$P_B^i = \max\left\{- P^i, \delta_{\min}h\eta_{dis}, ({b_{i-1} - b_{\max}}){\eta_{dis}}  \right\} $,\\
$\bullet$ If ${z}_i < 0$ then battery should charge such that 
	$P_B^i =\min$ $\left\{-P^i, \delta_{\max}h /\eta_{ch},
	{b_{i-1} - b_{\min}}/{\eta_{ch}}
	\right\} $.

\subsubsection{With C2 or C3 with NEM}
Optimal energy storage arbitrage under net-energy metering is introduced in \cite{hashmi2018netmetering}. The proposed algorithm provides a unique solution. However, performing arbitrage under time-of-use setting often has infinite possible optimal solutions as many different charging and discharging trajectories would lead to optimal arbitrage gains. 

A battery performing arbitrage under contracts C2 and C3 in Uruguay will perform at maximum 1 cycle over a day of depth-of discharge equal to $\text{SoC}_{\max}- \text{SoC}_{\min}$. Consider the the battery capacity denoted as $b_{\text{rated}}$ could charge from $\text{SoC}_{\min}$ to $\text{SoC}_{\max}$ at off-peak period then the storage buying cost including charging losses would be $\lambda_{\text{off-peak}} (\text{SoC}_{\max} - \text{SoC}_{\min}) b_{\text{rated}} / \eta_{\text{ch}}$. Similarly, for battery discharging from $\text{SoC}_{\max}$ to $\text{SoC}_{\min}$ during peak period of ToU price, $\lambda_{\text{peak}}$, would provide a revenue of $\lambda_{\text{peak}} (\text{SoC}_{\max} - \text{SoC}_{\min}) b_{\text{rated}} \eta_{\text{dis}}$. The storage owner profit is the difference of the revenue made by discharging and cost incurred during charging, shown as
\begin{equation}
G_{\text{arb}} = (\text{SoC}_{\max} - \text{SoC}_{\min}) b_{\text{rated}} \Big\{ \lambda_{\text{peak}} \eta_{\text{dis}} - \lambda_{\text{off-peak}} / \eta_{\text{ch}} \Big\}.
\end{equation}

Storage ramp rate selection under ToU is governed by the relationship between (a.) $T_{\text{off-peak}}$ and $T_{\text{ch}}$ and (b.) $T_{\text{peak}}$ and $T_{\text{dis}}$. For contract C2 we define $T_{\text{off-peak}}$ as the period of 17 hours (from 00:00 to 17:00) and $T_{\text{peak}}$ as the period of 6 hours (from 17:00 to 23:00). Refer to Fig.~\ref{c2nem}.

For contract C3 we define $T_{\text{off-peak}}$ as the period of 7 hours (from 00:00 to 7:00) and $T_{\text{peak}}$ as the period of 6 hours (from 17:00 to 23:00). Refer to Fig.~\ref{c3nem}.

Based on relationship between (a.) $T_{\text{off-peak}}$ and $T_{\text{ch}}$ and (b.) $T_{\text{peak}}$ and $T_{\text{dis}}$
following cases are proposed for storage performing arbitrage under ToU prices: 

 Case 1: $T_{\text{off-peak}} > T_{\text{ch}}$ and $T_{\text{peak}} > T_{\text{dis}}$: For this case the battery can be fully charged during off-peak period and fully discharged during peak period. In this case the ramp rate of the battery should be selected so the stress on battery (which is proportional to square of ramp rate) is minimized. The storage ramp rate is as follows:
	\begin{itemize}
		\item If \{$p_{elec}^i= \lambda_{\text{peak}}$\}  then discharge at $\delta_{\text{peak}}^* = \max ( (b_{\min} - b_{\max})/ T_{\text{peak}}, (b_{\min} - b_{i-1})/h, ~~\delta_{\min} )$,
		\item If \{$p_{elec}^i= \lambda_{\text{off-peak}}$\} then charge at $\delta_{\text{off-peak}}^* = \min ( (b_{\max} - b_{\min})/ T_{\text{off-peak}}, (b_{\max} - b_{i-1})/h, ~~\delta_{\max} )$,
	\end{itemize}
	
 Case 2: $T_{\text{off-peak}} > T_{\text{ch}}$ and $T_{\text{peak}} < T_{\text{dis}}$: Battery should only be charged to a level which could be discharged completely during peak period. The storage ramp rate is as follows:
	\begin{itemize}
		\item If \{$p_{elec}^i= \lambda_{\text{peak}}$\}  then discharge at $\delta_{\text{peak}}^* = \delta_{\min}$,
		\item If \{$p_{elec}^i= \lambda_{\text{off-peak}}$\} then charge at $\delta_{\text{off-peak}}^* = \min ( (b_{\max} - b_{\min})/ T_{\text{off-peak}}, (b_{\max} - b_{i-1})/h, ~~\delta_{\max} )$,
	\end{itemize}
	
 Case 3: $T_{\text{off-peak}} < T_{\text{ch}}$ and $T_{\text{peak}} > T_{\text{dis}}$: For this case battery cannot be charged completely from $b_{\min}$ to $b_{\max}$ within $T_{\text{off-peak}}$ time period. The storage ramp rate is as follows:
	\begin{itemize}
		\item If \{$p_{elec}^i= \lambda_{\text{peak}}$\}  then discharge at $\delta_{\text{peak}}^* = \max ( (b_{\min} - b_{\max})/ T_{\text{peak}}, (b_{\min} - b_{i-1})/h, ~~\delta_{\min} )$,
		\item If \{$p_{elec}^i= \lambda_{\text{off-peak}}$\} then charge at $\delta_{\text{off-peak}}^* = \delta_{\max}$,
	\end{itemize} 
	
 Case 4: $T_{\text{off-peak}} < T_{\text{ch}}$ and $T_{\text{peak}} < T_{\text{dis}}$: For this case the storage ramp rate is given by conditions as follows:
	\begin{itemize}
		\item If \{$p_{elec}^i= \lambda_{\text{peak}}$\}  then discharge at $\delta_{\text{peak}}^* = \delta_{\min}$,
		\item If \{$p_{elec}^i= \lambda_{\text{off-peak}}$\} then charge at $\delta_{\text{off-peak}}^* = \delta_{\max}$,
	\end{itemize}

It is clear that for Case 1 described above, the battery will perform 1 cycle per day at a Depth-of-Discharge equal to DoD~$=\frac{(b_{\max} - b_{\min})}{b_{\text{rated}}}= \text{SoC}_{\max} - \text{SoC}_{\min}$. Based on the degradation model proposed in \cite{hashmi2018long}, the number of cycles of operation could be controlled by increasing $\text{SoC}_{\min}$ and/or by decreasing $\text{SoC}_{\max}$. The control of cycles of operation using friction coefficient is introduced in \cite{hashmi2018limiting}.

We would like to highlight that similar to prior works \cite{hashmi2018limiting,hashmi2018long} for mid-peak period under contract C3, the optimal action is to do nothing (i.e. stay idle with $\delta_{\text{mid-peak}}^*=0$). 
Note that the proposed ramp rate thresholds presented also minimized the stress on the battery; stress as defined earlier on the battery is proportional to the square of the ramp rate  \cite{mathieu2016controlling, fortenbacher2017modeling}.

\subsection{Compensation Strategy for Reactive Power}
The unique attribute of contracts in Uruguay is the mechanism used for billing reactive power. The utility have two thresholds for charging for reactive power. First we list the penalty \\
$\bullet$ For PF $\in [0.92, 1]$: Consumes pay no penalty under contract C1,\\
$\bullet$ For PF $\in [0.82, 0.92]$: Consumes pay penalty under contract C1, C2 and C3,\\
$\bullet$ For PF $\in [0, 0.82]$: Consumes pay higher penalty compared to case where PF $\in [0.82, 0.92]$ under contract C1, C2 and C3,

Consumers could reduce there cost of consumption by maintaining a high PF for contracts C2 and C3:\\
$\bullet$ Maintaining a PF above 0.92 could provide additional gains to consumers opting for C2 and C3. However, consumers under C1 hold no incentive in improving the PF beyond 0.92.\\
$\bullet$ Improving PF which under nominal case is lower than 0.92 would reduce the cost of consumption under all contracts.

Next we describe the control mechanism for improving power factor or in other words reactive power compensation using energy storage.
\subsubsection{Reactive Power Compensation using Energy Storage interfaced via a Converter}
\label{reactivepfcmyop}
Authors in \cite{hashmi2018pfc} present that reactive compensation using energy storage interfaced via a converter performing arbitrage is largely decoupled. This is due to reactive power capability is governed by converter size and instantaneous active power. However, if the converter is slightly over-sized then converter could supply reactive power without being constrained primarily by instantaneous active power. Furthermore, since the converter size is static and a non-varying parameter therefore, lookahead in time is not required. This implies myopic reactive power compensation matches in optimal solution. Authors in \cite{hashmi2018pfc} show through numerical results that no lookahead reactive power compensation along with arbitrage (which requires lookahead) matches very closely with the co-optimization results where the storage performs arbitrage and PF compensation. The former optimization problem is denoted as $P_{rh}$ and the later optimization problem is denoted as $P_{plt}$ in \cite{hashmi2018pfc}. Based on these findings in prior work, we propose myopic reactive power compensation for all contracts. For users opting for C1, must not correct the PF beyond 0.92.

\section{Control Algorithm for Energy Storage for LV Consumers in Uruguay}
\label{sectionUrug5}
In previous sections we discussed the new contracts available for LV consumers in Uruguay and what roles energy storage could play for such consumers. In this section we present the storage control algorithm for the different contracts in Uruguay. The storage control algorithm consists of two algorithms: \textbf{\texttt{UruguayStorageControl}} and \textbf{\texttt{CalculateKfac}}.

These algorithms could be used for contracts C1, C2 or C3. The consumer needs to specify storage parameters, converter size, initial battery capacity. The code updates the storage control decision based on battery capacity available until the end of month is reached. At the end of month the algorithm calculates the profit due to storage integration for the consumer compared to the nominal case where no storage is considered.

\begin{algorithm*}
	\small{\textbf{Global Inputs}: Battery Characteristics: {$\eta_{\text{ch}}, \eta_{\text{dis}}, \delta_{\max}, \delta_{\min}, b_{\max}, b_{\min}$}}.\vspace{0.5mm}\\
	\small{\textbf{Inputs}: {Sampling time $h$, Number of points in a month $N_{\text{month}}$, Time instant index $i=0,$}} \vspace{0.5mm}\\ 
	\textbf{Function}: {Computes optimal Active and Reactive Power of Energy Storage Output for Contracts C1, C2 and C3}
	\begin{algorithmic}[1]
		\State Initialize $b_0$,
		and Input electricity prices $\lambda_{\text{peak}}, \lambda_{\text{mid-peak}}, \lambda_{\text{off-peak}}$,
		\State Input periods of peak, mid-peak and off-peak as $T_{\text{peak}}, T_{\text{mid-peak}}, T_{\text{off-peak}}$, \vspace{1.5mm}
		\State Set $E_a^{\text{peak}}, E_a^{\text{mid-peak}}, E_a^{\text{off-peak}},     E_{a+}^{\text{peak}}, E_{a+}^{\text{mid-peak}}, E_{a+}^{\text{off-peak}} =0$, Set $E_r, E_{rQ1} =0$,
		\State Input the converter rating $ S_B^{\max}$, $b_0$\vspace{1mm}
		\While{$i < N_{\text{month}}$}
		\State $i=i+1$, 
	\If{Contract is C1} Storage can only be used for increasing self-consumption as no Arbitrage possible
	\If{Net load without solar ${z}_i \geq 0$}
	Battery should discharge s.t. $x_i^* = \max\left\{- {z}_i/\eta_{dis}, \delta_{\min}h, ({b_{i-1} - b_{\max}}) \right\} $,
	\Else
~~~Battery should charge such that 	$x_i^* = \min\left\{-{z}_i \eta_{ch}, \delta_{\max}h , 	{b_{i-1} - b_{\min}}	\right\} $.
	\EndIf
	\ElsIf{Contract is C2 or C3}
		\If{$p_{elec}^i== \lambda_{\text{peak}}$} 
~~Battery should discharge, $x_i^* = h \max \Big( (b_{\min} - b_{\max})/ T_{\text{peak}}, (b_{\min} - b_0)/h, ~~\delta_{\min} \Big)$,
		\ElsIf{$p_{elec}^i== \lambda_{\text{mid-peak}}$}
~~Battery should do nothing, $x_i^* = 0$,
		\ElsIf{$p_{elec}^i== \lambda_{\text{off-peak}}$}
		Battery should charge, $x_i^* = h \min \Big( (b_{\max} - b_{\min})/ T_{\text{off-peak}}, (b_{\max} - b_0)/h, ~~\delta_{\max} \Big)$,
		\EndIf 
	\EndIf
		\State $s_i^* = [x_i^*]^+/\eta_{ch} - [x_i^*]^-\eta_{dis}$ and  Set $P_B^i = sign(s_i^*) \times \min\Big(|s_i^*|, S_B^{\max}\Big)$,
		\If{Contract is C2 and C3}
		~~ Set $Q_B^i = -sign(Q^i)~~\times ~ \min \Big(|{Q}^i|, ~~ \sqrt{(S_B^{\max})^2 - (s_i^*)^2}\Big) $,
		\Else ~ Select $Q_B^i$ such that power factor is no more then 0.92, as C1 consumers have no additional incentive
		\EndIf
		 \vspace{2mm}
		\If{$p_{elec}^i== \lambda_{\text{peak}}$} 
		Calculate  $E_a^{\text{peak}} = E_a^{\text{peak}} + (P^i + P_B^i)h$, and calculate $(E_a^{\text{peak}})_{\text{nominal}} = (E_a^{\text{peak}})_{\text{nominal}} + (P^i)h$, \vspace{1.5mm}
		\ElsIf{$p_{elec}^i== \lambda_{\text{mid-peak}}$}
 Calculate $E_a^{\text{mid-peak}} = E_a^{\text{mid-peak}} + (P^i + P_B^i)h$, and calculate $(E_a^{\text{mid-peak}})_{\text{nominal}} = (E_a^{\text{mid-peak}})_{\text{nominal}} + (P^i)h$, \vspace{0.5mm}
		\ElsIf{$p_{elec}^i== \lambda_{\text{off-peak}}$}
Calculate$E_a^{\text{off-peak}} = E_a^{\text{off-peak}} + (P^i + P_B^i)h$, and calculate $(E_a^{\text{off-peak}})_{\text{nominal}} = (E_a^{\text{off-peak}})_{\text{nominal}} + (P^i)h$,
		\EndIf \vspace{2mm}
		\State Aggregate reactive power,~~ $E_r = E_r + (Q^i + Q_B^i)h$,
		\State Aggregate reactive power in Quadrant 1,   ~~ $E_{rQ1} = E_{rQ1} + |(Q^i + Q_B^i)h|$, \vspace{1.5mm}
		\State $b_i^*= b_{0}+x_i^* $ and Update $b_0=b_i^*$, \vspace{1.5mm}
		\EndWhile \vspace{1mm}
		 \vspace{1mm}
		\If{Contract is C1} 
		\State Calculate $(K_{\text{fac}})_{\text{nominal}}$ with C1 contract using \textbf{\texttt{CalculateKfac}} defined as Algorithm~\ref{alg:3},
		\State Calculate $K_{\text{fac}}$ with C1 contract using \textbf{\texttt{CalculateKfac}},
		\State Nominal Cost  $=\Big(\lambda_{\text{fixed}} + (K_{\text{fac}})_{\text{nominal}}\Big) (E_a)_{\text{nominal}}$,
		\State New Cost $=\Big(\lambda_{\text{fixed}} + K_{\text{fac}}\Big) E_a$, \vspace{1mm}

		\ElsIf{Contract is C2} 
		\State Calculate $(K_{\text{fac}})_{\text{nominal}}$ with C2 contract using \textbf{\texttt{CalculateKfac}} defined as Algorithm~\ref{alg:3},
		\State Calculate $K_{\text{fac}}$ with C2 contract using \textbf{\texttt{CalculateKfac}},
		\State Nominal Cost  $=\Big(\lambda_{\text{peak}} + (K_{\text{fac}})_{\text{nominal}}\Big) (E_a^{\text{peak}})_{\text{nominal}} +  \lambda_{\text{off-peak}} \Big((E_a^{\text{mid-peak}})_{\text{nominal}} + (E_a^{\text{off-peak}})_{\text{nominal}}\Big)$,\vspace{0.5mm}
		\State New Cost $=\Big(\lambda_{\text{peak}} + K_{\text{fac}}\Big) E_a^{\text{peak}} +  \lambda_{\text{off-peak}} \Big(E_a^{\text{mid-peak}} + E_a^{\text{off-peak}}\Big)$, \vspace{1mm}
		\ElsIf{Contract is C3}
		\State Calculate $E_{a+}^{\text{peak}} = |E_{a}^{\text{peak}}|$,    ~~~~~~ $E_{a+}^{\text{mid-peak}} = | E_{a}^{\text{mid-peak}}|$,    ~~~~~~ $E_{a+}^{\text{off-peak}} = |E_{a}^{\text{off-peak}}|$,
		\State Calculate $(K_{\text{fac}})_{\text{nominal}}$ with C3 contract using \textbf{\texttt{CalculateKfac}},
		\State Calculate $K_{\text{fac}}$ with C3 contract using \textbf{\texttt{CalculateKfac}},
		\State Nominal Cost $=\lambda_{\text{peak}} (E_a^{\text{peak}})_{\text{nominal}} +   \lambda_{\text{mid-peak}}(E_a^{\text{mid-peak}})_{\text{nominal}} + \lambda_{\text{off-peak}}(E_a^{\text{off-peak}})_{\text{nominal}} + (K_{\text{fac}})_{\text{nominal}}(E_{a+}^{\text{peak}})_{\text{nominal}}$,
		\State New Cost $=\lambda_{\text{peak}} E_a^{\text{peak}} +   \lambda_{\text{mid-peak}}E_a^{\text{mid-peak}} + \lambda_{\text{off-peak}}E_a^{\text{off-peak}} + K_{\text{fac}}E_{a+}^{\text{peak}}$,
		\EndIf 
		\State Profit = Nominal Cost of Consumption $-$ New Cost of Consumption with inclusion Storage,
		\State Return Vectors $x^*, Q_B$ and Profit.
	\end{algorithmic}
	\caption{\textbf{\texttt{UruguayStorageControl}}}\label{alg:2}
\end{algorithm*}

\begin{algorithm*}
	\small{\textbf{Inputs}: {$E_a^{\text{peak}}, E_a^{\text{mid-peak}}, E_a^{\text{off-peak}}, (E_a^{\text{peak}})_{\text{nominal}}, (E_a^{\text{mid-peak}})_{\text{nominal}}, (E_a^{\text{off-peak}})_{\text{nominal}}, E_r, E_{rQ1}$}} \vspace{1.5mm}\\ 
	\textbf{Function}: {Computes $K_{\text{fac}}$ for Contracts C1, C2 and C3 for Uruguay LV consumers}\\
	\begin{algorithmic}[1]
		\State Initialize $A, B$, \vspace{2mm}
		\If{Contract is C1} 
		\State Calculate $E_a =E_a^{\text{peak}} +E_a^{\text{mid-peak}}+E_a^{\text{off-peak}}$,
		\State Calculate $(E_a) =(E_a^{\text{peak}})_{\text{nominal}} +(E_a^{\text{mid-peak}})_{\text{nominal}}+(E_a^{\text{off-peak}})_{\text{nominal}}$, \vspace{2mm}
		\If{$E_r/E_a \leq 0.426$}
		\State Assign $K_{\text{fac}} = 0,$
		\ElsIf{$0.426 < E_r/E_a \leq 0.7$}
		\State Assign $K_{\text{fac}} = 0.4 \left(\frac{E_r}{E_a} - 0.426 \right),$
		\Else
		\State Assign  $K_{\text{fac}} =   0.4 \left(\frac{E_r}{E_a} - 0.426 \right) + 0.6 \left(\frac{E_r}{E_a} - 0.7 \right) $,
		\EndIf
		\If{$(E_r)_{\text{nominal}}/(E_a)_{\text{nominal}} \leq 0.426$}
		\State Assign $(K_{\text{fac}})_{\text{nominal}} = 0,$
		\ElsIf{$0.426 < (E_r)_{\text{nominal}}/(E_a)_{\text{nominal}} \leq 0.7$}
		\State Assign $(K_{\text{fac}})_{\text{nominal}} = 0.4 \left(\frac{(E_r)_{\text{nominal}}}{(E_a)_{\text{nominal}}} - 0.426 \right),$
		\Else
		\State Assign  $(K_{\text{fac}})_{\text{nominal}} =   0.4 \left(\frac{(E_r)_{\text{nominal}}}{(E_a)_{\text{nominal}}} - 0.426 \right) + 0.6 \left(\frac{(E_r)_{\text{nominal}}}{(E_a)_{\text{nominal}}} - 0.7 \right) $,
		\EndIf \vspace{3mm}

		\ElsIf{Contract is C2} 
		\State Calculate $E_a =E_a^{\text{peak}} +E_a^{\text{mid-peak}}+E_a^{\text{off-peak}}$,
		\State Calculate $(E_a)_{\text{nominal}} =(E_a^{\text{peak}})_{\text{nominal}} +(E_a^{\text{mid-peak}})_{\text{nominal}}+(E_a^{\text{off-peak}})_{\text{nominal}}$,\vspace{2mm}
		\If{$E_r/E_a \leq 0.7$}
		\State Assign $K_{\text{fac}} = \frac{B}{100} \left(\frac{E_r}{E_a} - 0.426 \right),$
		\Else
		\State Assign  $K_{\text{fac}} =   \frac{B}{100} \left(\frac{E_r}{E_a} - 0.426 \right) + \frac{100-B}{100} \left(\frac{E_r}{E_a} - 0.7 \right) $,
		\EndIf
		
		\If{$(E_r)_{\text{nominal}}/(E_a)_{\text{nominal}} \leq 0.7$}
		\State Assign $(K_{\text{fac}})_{\text{nominal}} = \frac{B}{100} \left(\frac{(E_r)_{\text{nominal}}}{(E_a)_{\text{nominal}}} - 0.426 \right),$
		\Else
		\State Assign  $(K_{\text{fac}})_{\text{nominal}} =   \frac{B}{100} \left(\frac{(E_r)_{\text{nominal}}}{(E_a)_{\text{nominal}}} - 0.426 \right) + \frac{100-B}{100} \left(\frac{(E_r)_{\text{nominal}}}{(E_a)_{\text{nominal}}} - 0.7 \right) $,
		\EndIf \vspace{3mm}

		\ElsIf{Contract is C3}
		\State Calculate $E_{a+}^{\text{peak}} = |E_{a}^{\text{peak}}|$,    ~~~~~~ $E_{a+}^{\text{mid-peak}} = | E_{a}^{\text{mid-peak}}|$,    ~~~~~~ $E_{a+}^{\text{off-peak}} = |E_{a}^{\text{off-peak}}|$,
		
				\State Calculate $E_{a+} =E_{a+}^{\text{peak}} +E_{a+}^{\text{mid-peak}}+E_{a+}^{\text{off-peak}}$,
				\State Calculate $(E_{a+})_{\text{nominal}} =(E_{a+}^{\text{peak}})_{\text{nominal}} +(E_{a+}^{\text{mid-peak}})_{\text{nominal}}+(E_{a+}^{\text{off-peak}})_{\text{nominal}}$, \vspace{2mm}
				\If{$E_{rQ1}/E_a \leq 0.7$}
				\State Assign $K_{\text{fac}} = \frac{A}{100} \left(\frac{E_{rQ1}}{E_{a+}} - 0.426 \right),$
				\Else
				\State Assign  $K_{\text{fac}} =   \frac{A}{100} \left(\frac{E_{rQ1}}{E_a} - 0.426 \right) + \frac{100-A}{100} \left(\frac{E_{rQ1}}{E_{a+}} - 0.7 \right) $,
				\EndIf
				
				\If{$(E_{rQ1})_{\text{nominal}}/(E_{a+})_{\text{nominal}} \leq 0.7$}
				\State Assign $(K_{\text{fac}})_{\text{nominal}} = \frac{A}{100} \left(\frac{(E_{rQ1})_{\text{nominal}}}{(E_{a+})_{\text{nominal}}} - 0.426 \right),$
				\Else
				\State Assign  $(K_{\text{fac}})_{\text{nominal}} =   \frac{A}{100} \left(\frac{(E_{rQ1})_{\text{nominal}}}{(E_{a+})_{\text{nominal}}} - 0.426 \right) + \frac{100-A}{100} \left(\frac{(E_{rQ1})_{\text{nominal}}}{(E_{a+})_{\text{nominal}}} - 0.7 \right) $,
				\EndIf \vspace{3mm}

		\EndIf \vspace{2mm}
		\State Return $(K_{\text{fac}})_{\text{nominal}}$, $K_{\text{fac}}$.
	\end{algorithmic}
	\caption{\textbf{\texttt{CalculateKfac}}}\label{alg:3}\vspace{2.5mm}
\end{algorithm*}

\subsection{Storage Operation Immune to Uncertainty}
\textit{  Uncertainty in Active Power Control in Uruguay:} 
Control of energy storage is coupled in time. For the active power, if the battery is charged in the present time then the amount of energy available in subsequent time instant will be higher. Authors in \cite{hashmi2018netmetering} identify that storage performing only active power arbitrage have uncertainty due to two parameters: (a.) uncertainty due to electricity price variation and (b.) uncertainty due net load without storage variation. These sources of uncertainty was for the case of real-time electricity price and where the selling price have an arbitrary ratio varying between 0 and 1 with respect to the buying price \cite{hashmi2018netmetering}, refer to Case 4 in Figure~\ref{arbitrageanduncertainty}. However, for the case of Uruguay, the LV consumers respond to Time-of-Use prices which vary in a deterministic manner, thus no uncertainty due to electricity price variations.

Prior work on energy arbitrage \cite{hashmi2017optimal} have identified that energy storage control under equal instantaneous buy and sell price becomes independent of the net load without storage. 
This implies that active power control faces no uncertainty with respect to contracts in Uruguay which have ToU pricing structure and NEM which provides equal buy and sell price at each time instant. Refer to Appendix~\ref{arbitragesec} and Figure~\ref{arbitrageanduncertainty} for details.

\textit{Uncertainty in Reactive Power Control in Uruguay:} It is made clear in Section~\ref{reactivepfcmyop} that storage interfaced via a converter operating with no lookahead reaches close to optimality. Furthermore, we show in numerical results that valuation of reactive power compensation is significantly lower than incentive in performing arbitrage. 
Therefore, in the algorithm presented for active and reactive energy compensation, the active power is adjusted to maximize arbitrage gains and the available energy storage converter is utilized for reactive power compensation, thus prioritizing active power over reactive power compensation.


%


\section{Numerical Experiments}
\label{sectionUrug6}
The numerical results presents three scenarios. In Section~\ref{onlyarbitragepotential} the potential of performing energy arbitrage is identified based on gains made by performing arbitrage for a month. In Section~\ref{nominalvswithstorage} consumer cost of consumption is compared with nominal case with storage to that of consumer load with energy storage performing active and/or reactive energy compensation. Based on 2 standard models of Tesla PowerWall we recommend the suitable contracts. A similar analysis would be required with consumer of different consumption pattern. In Section~\ref{profitabilitystorage} we present that based on storage operational cycles and degradation, the battery under certain contracts and size could be profitable for consumers. We observe the financial returns of storage in Uruguay are significantly higher than many ISOs in the USA and Europe \cite{hashmi2017optimal}. 
\subsection{Arbitrage Potential}
\label{onlyarbitragepotential}
In Table~\ref{arbitragepoten} we list the arbitrage gains the storage owners would make for $\text{SoC}_{\max} = 0.98$, $\text{SoC}_{\min} = 0.2$, $\eta_{\text{dis}}$=$\eta_{\text{ch}}$ =0.95.
		\begin{table}
			\caption {\small{Arbitrage Gains Potential for a month}}
			\label{arbitragepoten}
			\vspace{-10pt}
			\begin{center}
				\begin{tabular}{| c | c| c |}
					\hline
					$b_{\text{rated}}$ kWh& Contract C2 & Contract C3\\ 
					\hline
					\hline
					1&	106.68& 	147.32\\
					2&	213.36& 	294.65\\
					5&	533.40& 	736.62\\
					10&	1066.81&	1473.23\\
					20&	2133.62&	2946.46\\
					\hline
				\end{tabular}
				\hfill\
			\end{center}
		\end{table}
From Table~\ref{arbitragepoten} it is clear that storage is hugely profitable. Storage owner would make 3.56 peso per day per kWh for contract C2 and 4.91 peso per day per kWh for contract C3.
For a user with a 1000 peso per month as electricity bill under C3 would have to pay nothing perpetually if they install a 7kWh battery (performing arbitrage only) which could charge completely within the off-peak time period and discharge completely during peak periods.
\subsection{Consumer gains with/without storage}
\label{nominalvswithstorage}
For this numerical experiment we assume a consumer with active energy consumed over a month listed in the table ~\ref{activeenergy}.
		\begin{table}
			\caption {\small{Nominal load of a LV consumer}}
			\label{activeenergy}
			\vspace{-10pt}
			\begin{center}
				\begin{tabular}{| c | c| }
					\hline
					Load Consumed during& kWh\\ 
					\hline
					\hline
					Peak Period& 	200\\
					Mid Peak& 	200\\
					Off-peak&	100\\
					\hline
				\end{tabular}
				\hfill\
			\end{center}
		\end{table}
We assume the cumulative absolute value of reactive power is $E_{rQ1} = 1.2 \times E_r$ and cumulative value of absolute of active power $E_{a+}= E_a $. For the consumer load listed in Table~\ref{activeenergy} we vary the amount of reactive load and see the effect without and with inclusion of energy storage providing active and/or reactive energy compensation.

\subsubsection{Nominal Case}
The nominal case consists of inelastic load with DG output in absence of energy storage.
Thus the nominal case is described for no compensation of active and reactive energy at consumer end. It is essential to analyze the effect on consumer load listed in Table~\ref{activeenergy} on the cost of consumption based on the different contracts. Note the presented analysis will vary with consumer load and authors wish to present the mechanism of analysis for the case described, of course a similar analysis needs to be performed for a specific user who wish to select the appropriate contract based on their consumption behavior.
\begin{figure}
	\centering
	\includegraphics[width=3.5in]{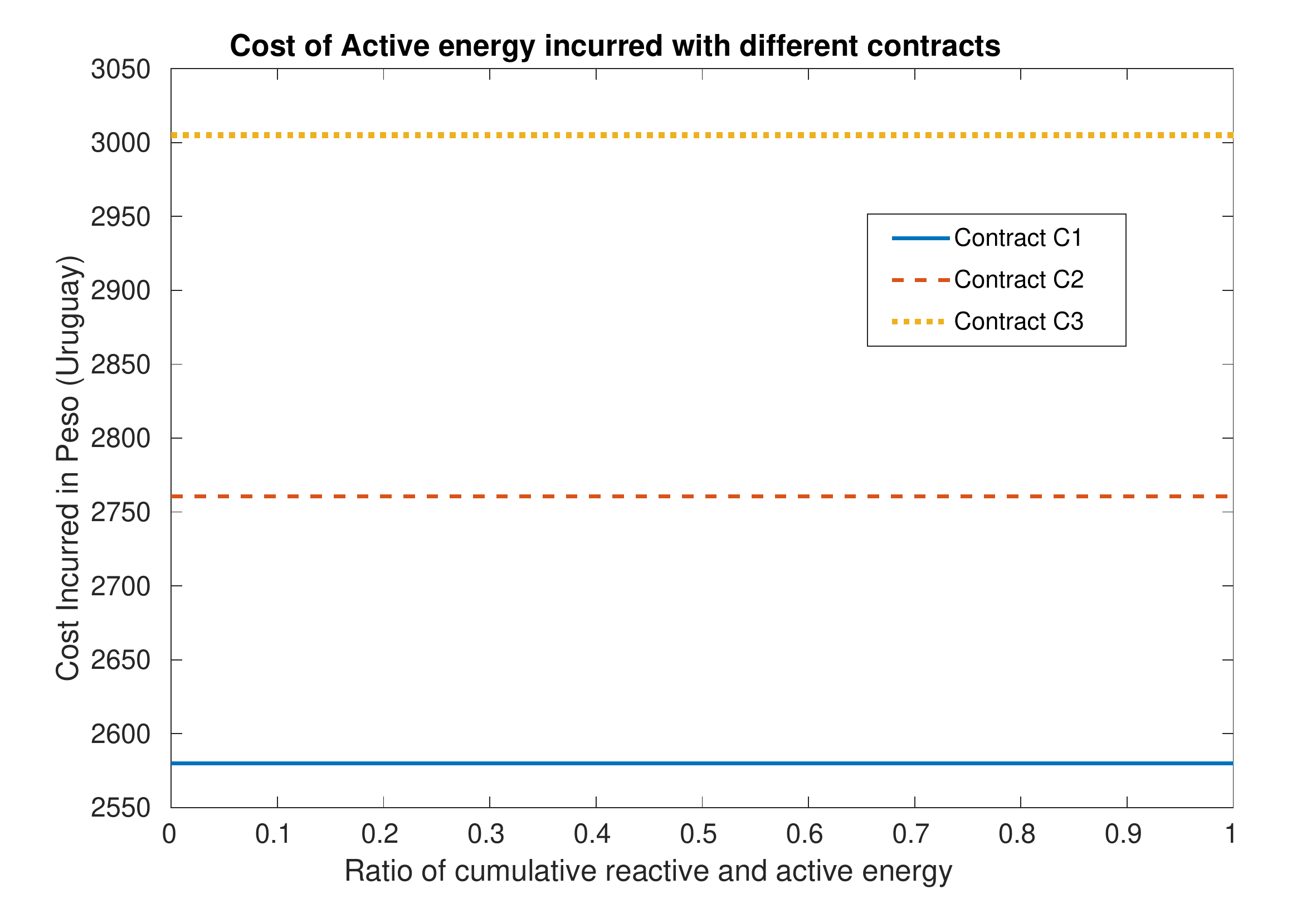}
	\caption{Active energy cost with different contracts}\label{res1}
\end{figure}

Fig.~\ref{res1} presents the the cost of active energy with increasing share of reactive energy. It is clear that cost of active energy is independent of increase in reactive energy. In absence of flexibility to alter the active energy consumption, clearly contract C1 is best suited for the consumer presented in Table~\ref{activeenergy}.
\begin{figure}
	\centering
	\includegraphics[width=3.2in]{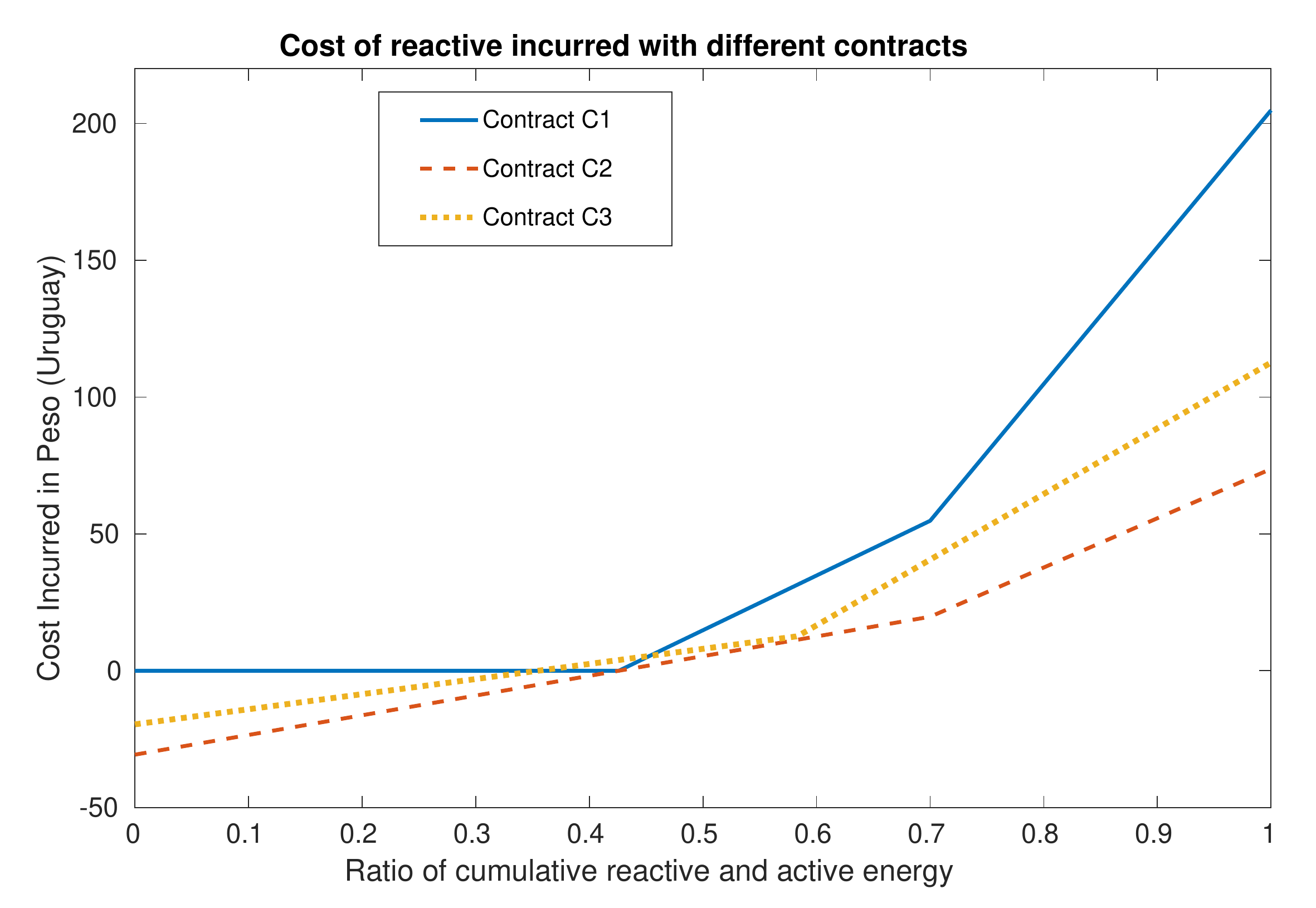}
	\caption{Reactive energy cost with different contracts}\label{res2}
\end{figure}
Fig.~\ref{res2} presents the cost of reactive energy under contract C1, C2 and C3. With increase in share of reactive energy, cost paid by consumer under C1 is significantly higher compared to C2 and C3.

\begin{figure}
	\centering
	\includegraphics[width=3.2in]{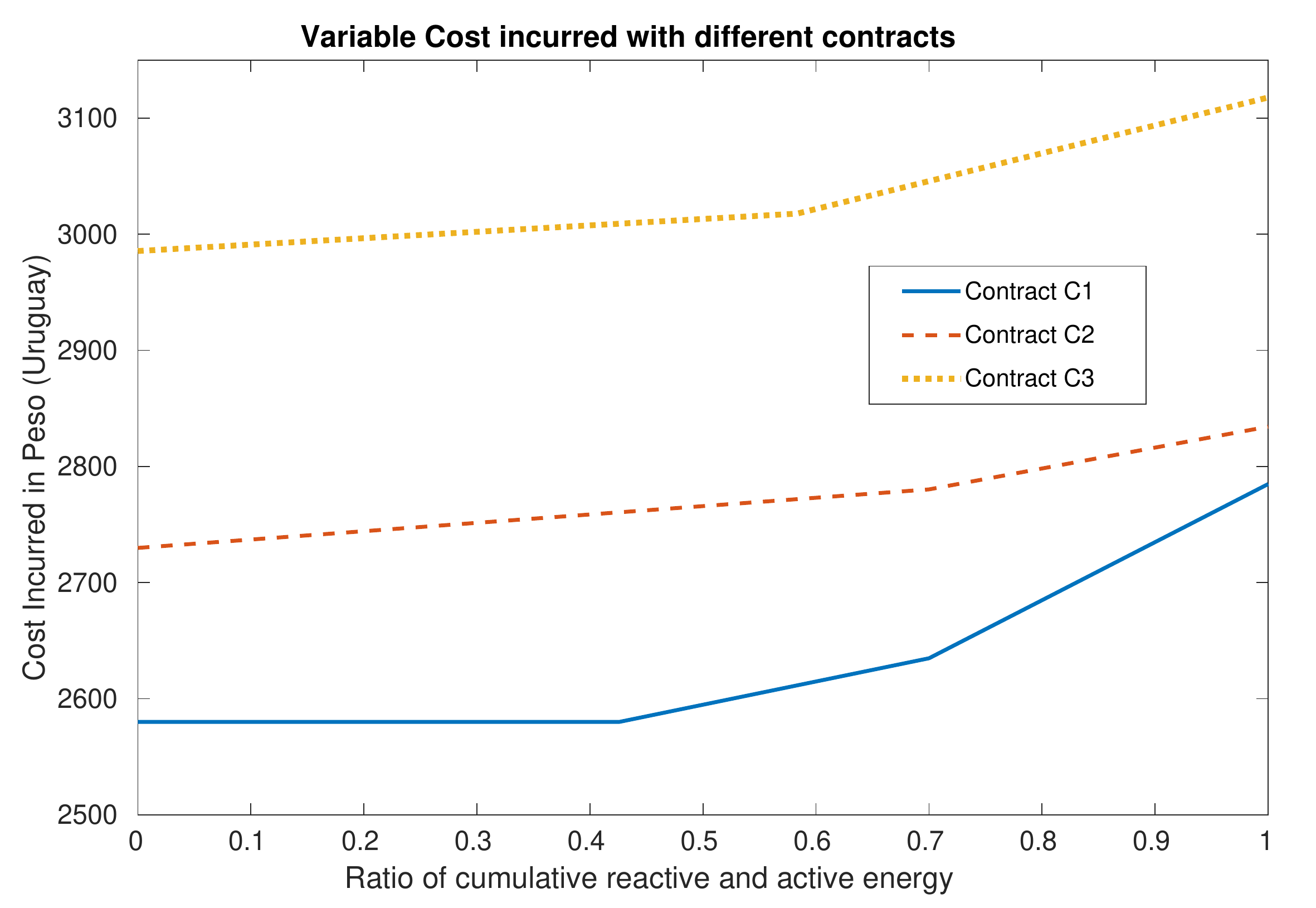}
	\caption{Variable cost with different contracts}\label{res3}
\end{figure}

\begin{figure}
	\centering
	\includegraphics[width=3.2in]{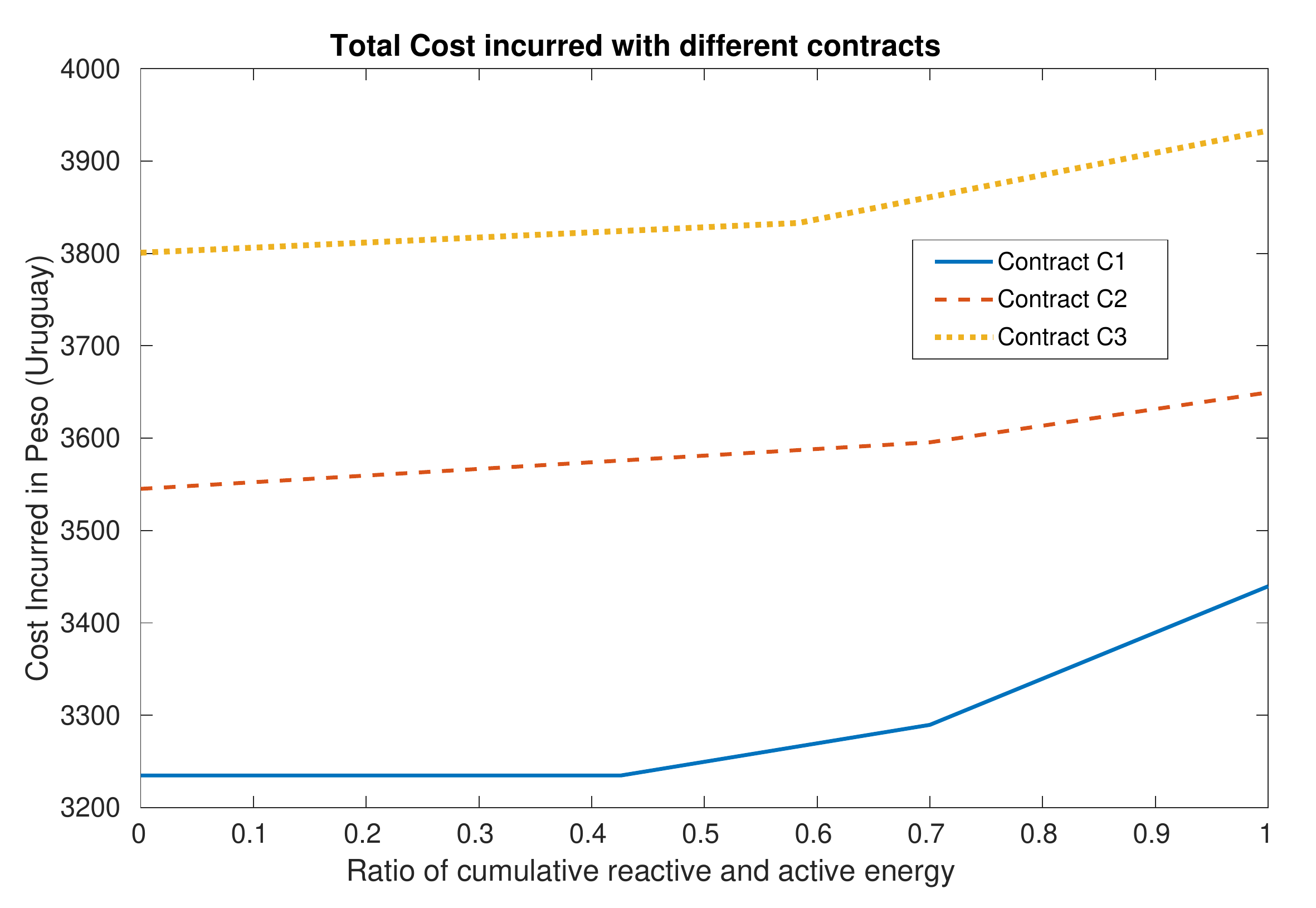}
	\caption{Total energy cost with different contracts}\label{res5}
\end{figure}

Fig.~\ref{res3} plots the variable cost component denoted as $C_{\text{variable}}^{Ci}$. It is clear that the cost of active power dominates the variable cost.
The same trends are visible in Fig.~\ref{res5} which presents the total cost of consumption which includes the variable cost, fixed cost and power cost. Note that the fixed cost for C2 and C3 is almost double the cost paid by consumer under C1.
\begin{figure}
	\centering
	\includegraphics[width=3.2in]{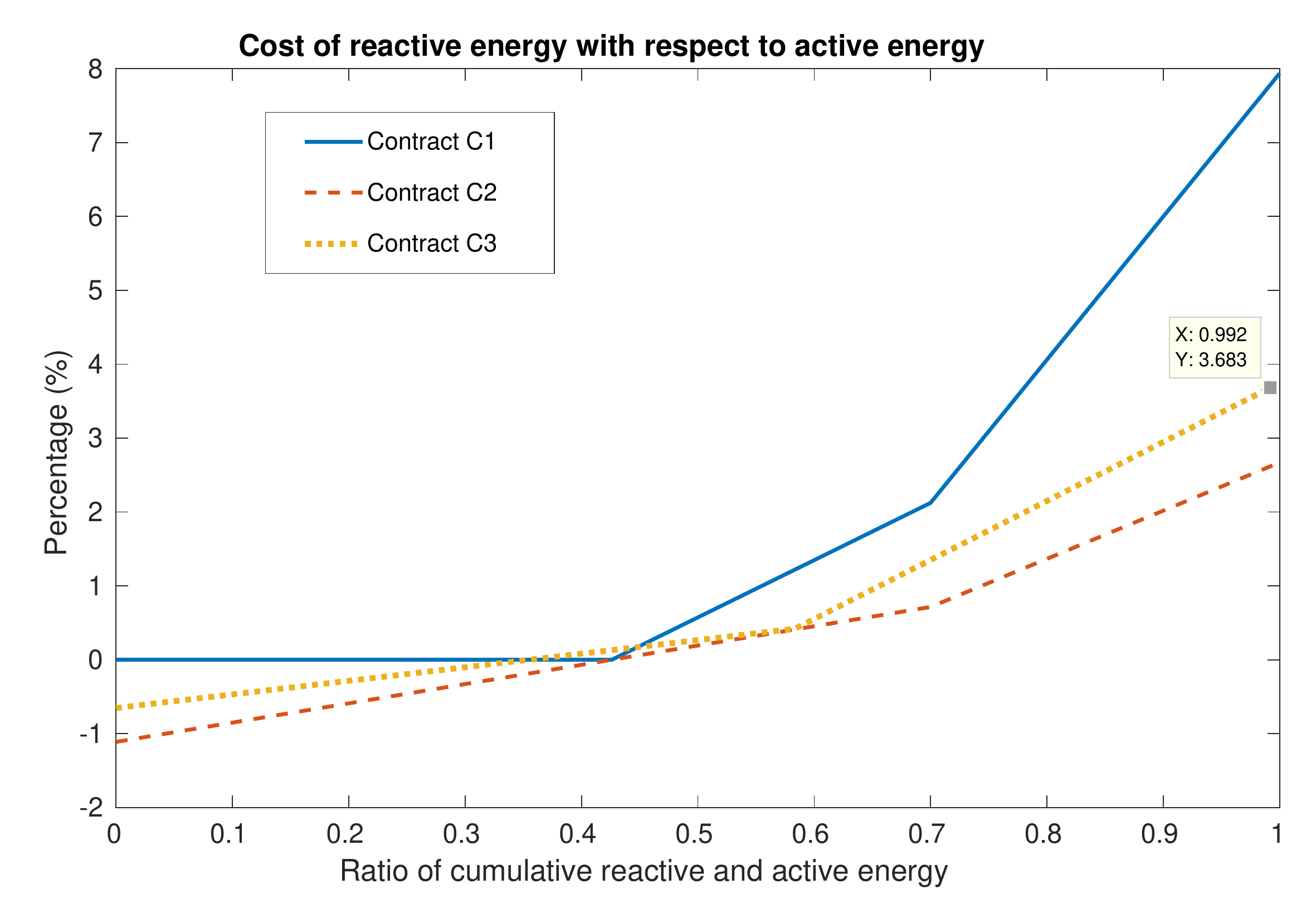}
	\caption{Cost of reactive energy compared to active energy cost with different contracts}\label{res4}
\end{figure}
Fig.~\ref{res4} presents the ratio of the cost of reactive energy and the cost of active energy in percentage. For consumers consuming equal quantities active and reactive energy end up paying almost 8\% of active energy cost under C1, less than 4\% of active energy cost under C2 and less than 3\% of active energy cost under C3. Clearly, the value of reactive energy compensation is highest under contract C1. We would like to bring to notice that the cost paid by consumer for active energy is significantly higher than that of the reactive energy. 
In the proposed algorithm the energy storage converter active power output is selected based on maximization of arbitrage gains and remaining converter capacity is used for supplying reactive power for correcting the PF in accordance to the contract selected by the LV consumer.

\subsubsection{Inclusion of Tesla PowerWall 1 (6.4 kWh)}
\begin{figure}
	\centering
	\includegraphics[width=3.2in]{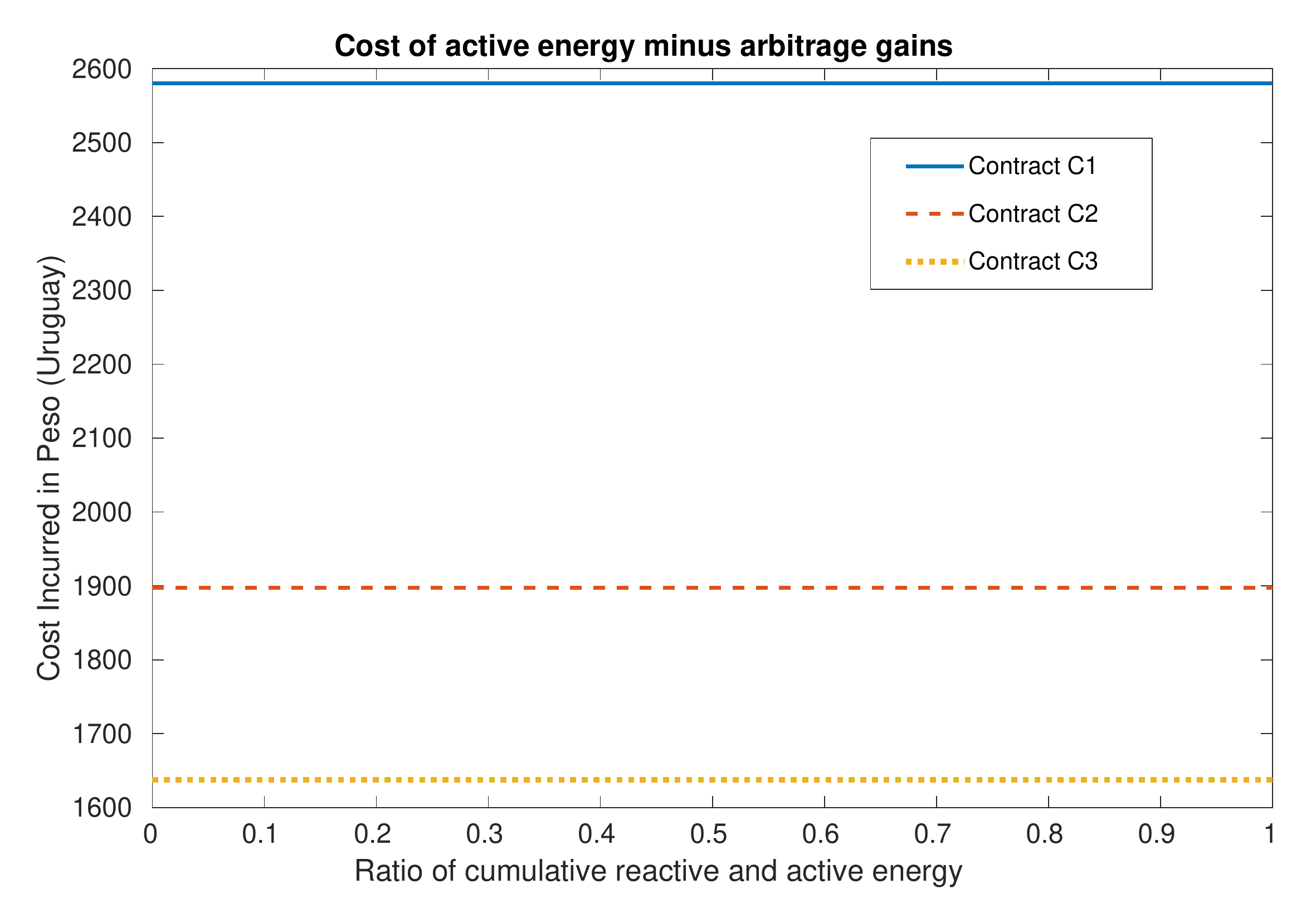}
	\caption{Cost of active energy with arbitrage}\label{res6}
\end{figure}

The total arbitrage gains made by consumer owning Tesla PowerWall 1 which has a rated capacity of 6.4 kWh is 
682.4733 pesos under contract C2 and 942.5827 pesos under contract C3. 
Fig.~\ref{res6} presents the cost of active energy with inclusion of energy storage performing only arbitrage. Now we can see that inclusion of storage have turned contract C3 most profitable for the consumer.

\begin{figure}
	\centering
	\includegraphics[width=3.2in]{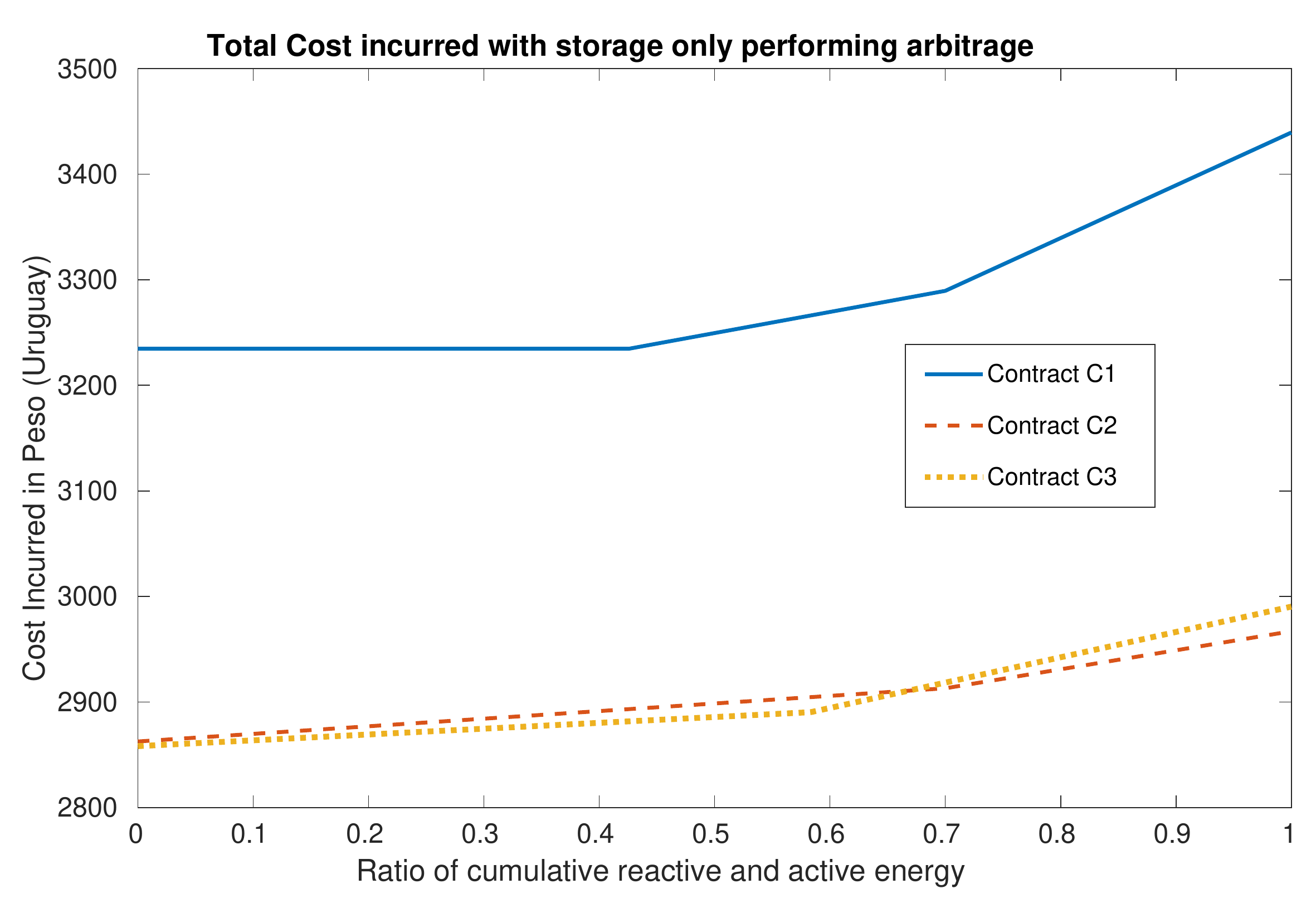}
	\caption{Total cost with only arbitrage}\label{res7}
\end{figure}

\begin{figure}
	\centering
	\includegraphics[width=3.2in]{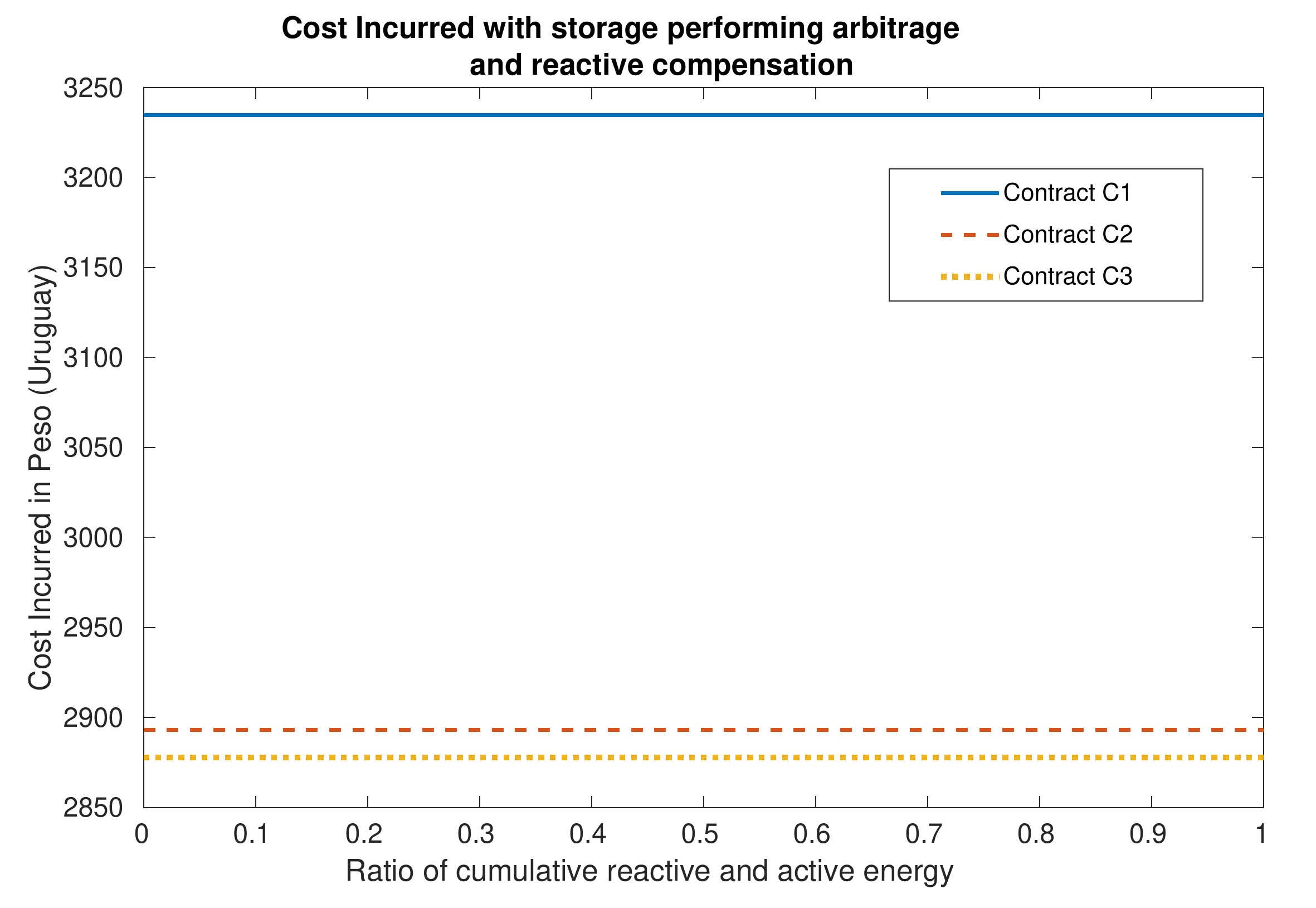}
	\caption{Total cost of electricity}\label{res8}
\end{figure}

\begin{figure}
	\centering
	\includegraphics[width=3.2in]{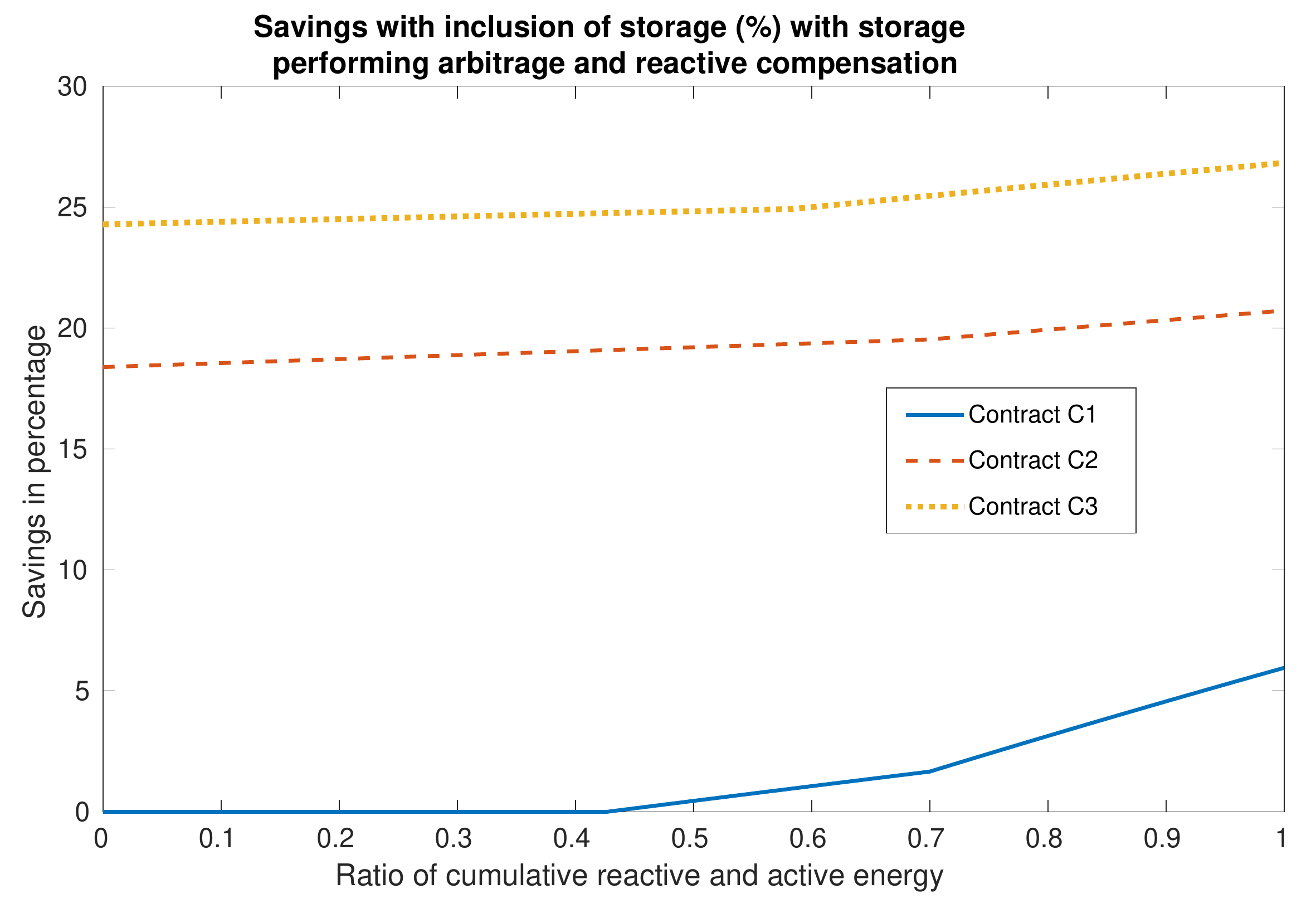}
	\caption{Profit due to inclusion of storage in terms of savings in percentage}\label{res9}
\end{figure}

Fig.~\ref{res7} presents the total cost with inclusion of energy storage performing only arbitrage. In this case, total cost under C2 and C3 are comparable.
Assuming energy storage converter could supply all the reactive energy that the user consumed, which is fairly reasonable as the utility aggregates reactive power over the month. Fig.~\ref{res8} presents the total cost of consumption (includes fixed and power cost) with energy storage performing arbitrage and reactive power compensation. Clearly, contract C3 consumer pays the least and consumer under C1 pays the maximum electricity bill in this scenario.
Fig.~\ref{res9} shows the percentage of consumer would make with respect to the nominal case under the same contract. Consumers with C3 would make a profit exceeding 25\% by installing Tesla PowerWall 1. Note since consumer under C1 could perform only reactive power compensation, it makes a profit only if the ratio of reactive energy and active energy exceeds 0.426.

\subsubsection{Inclusion of Tesla PowerWall 2 (13.5 kWh)}
The total arbitrage gains made by consumer owning Tesla PowerWall 2 which has a rated capacity of 13.5 kWh is 
1439.6 pesos under contract C2 and 1988.3 pesos under contract C3. 

\begin{figure}
	\centering
	\includegraphics[width=3.2in]{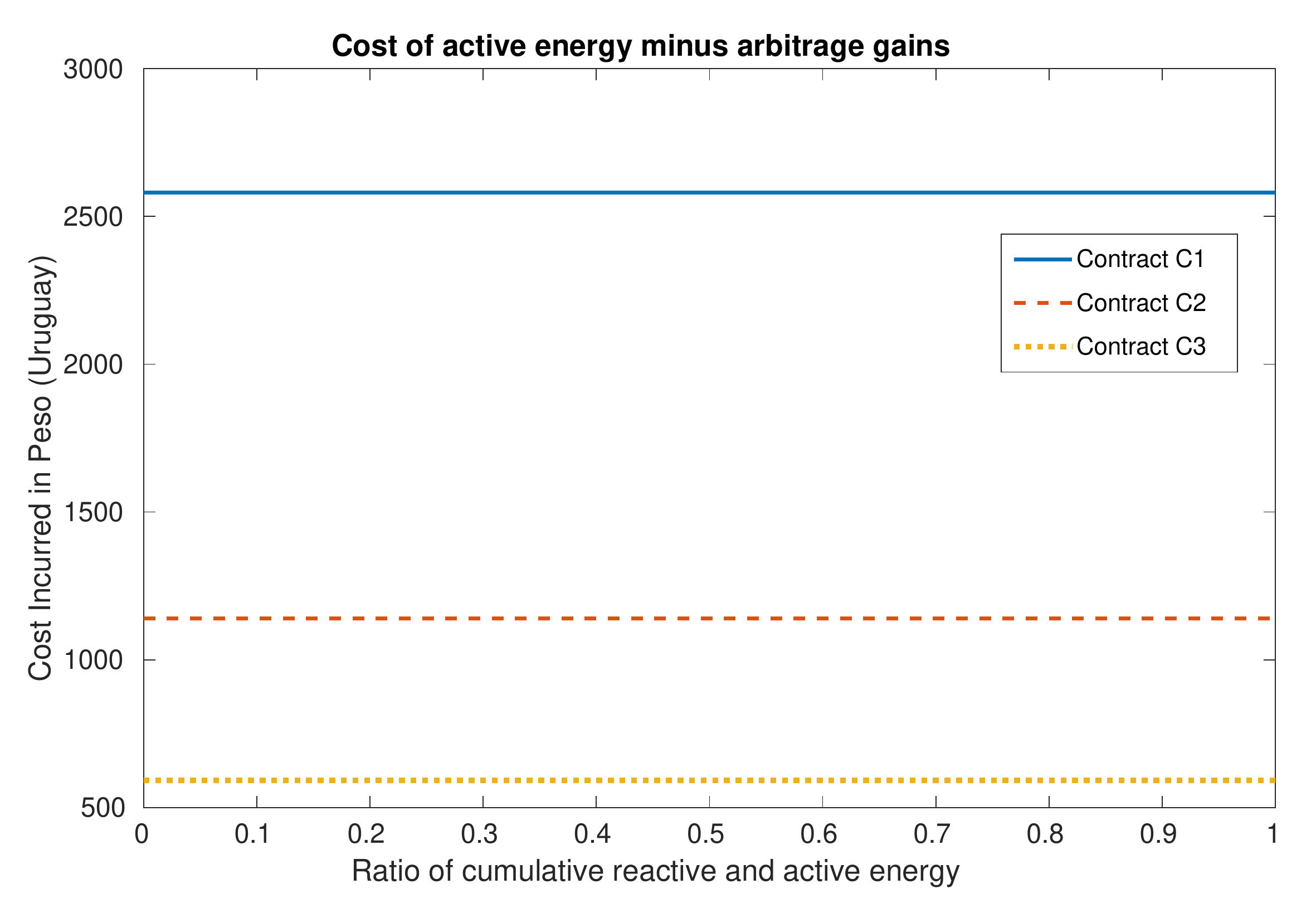}
	\caption{Active energy cost with arbitrage for PowerWall 2}\label{res10}
\end{figure}

Fig.~\ref{res10} presents the cost of active energy with inclusion of energy storage performing only arbitrage. Now we can see that inclusion of storage have turned contract C3 most profitable for the consumer. Note that the gap between C2 and C3 increases as the storage size increases, as the amount of profit under C3 is higher for performing arbitrage compared to C2. Compare Fig.~\ref{res6} and Fig.~\ref{res10}.

\begin{figure}
	\centering
	\includegraphics[width=3.2in]{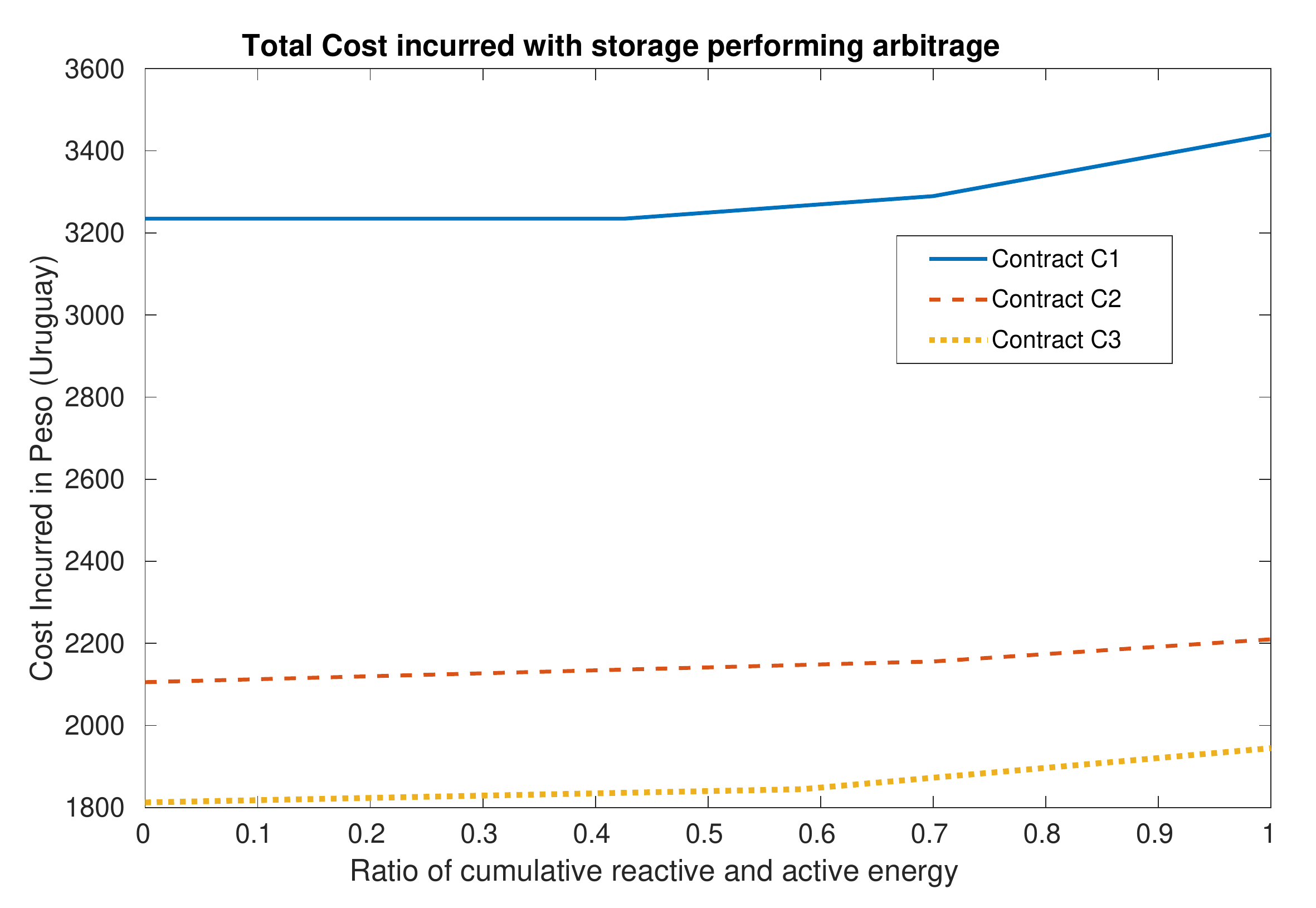}
	\caption{Total cost with only arbitrage for PowerWall 2}\label{res11}
\end{figure}

Fig.~\ref{res11} presents the total cost with inclusion of energy storage performing only arbitrage. In this case, total cost under C2 and C3 are not comparable. Consumer would pay around 150 pesos lower under C3 then under C2.

\begin{figure}
	\centering
	\includegraphics[width=3.2in]{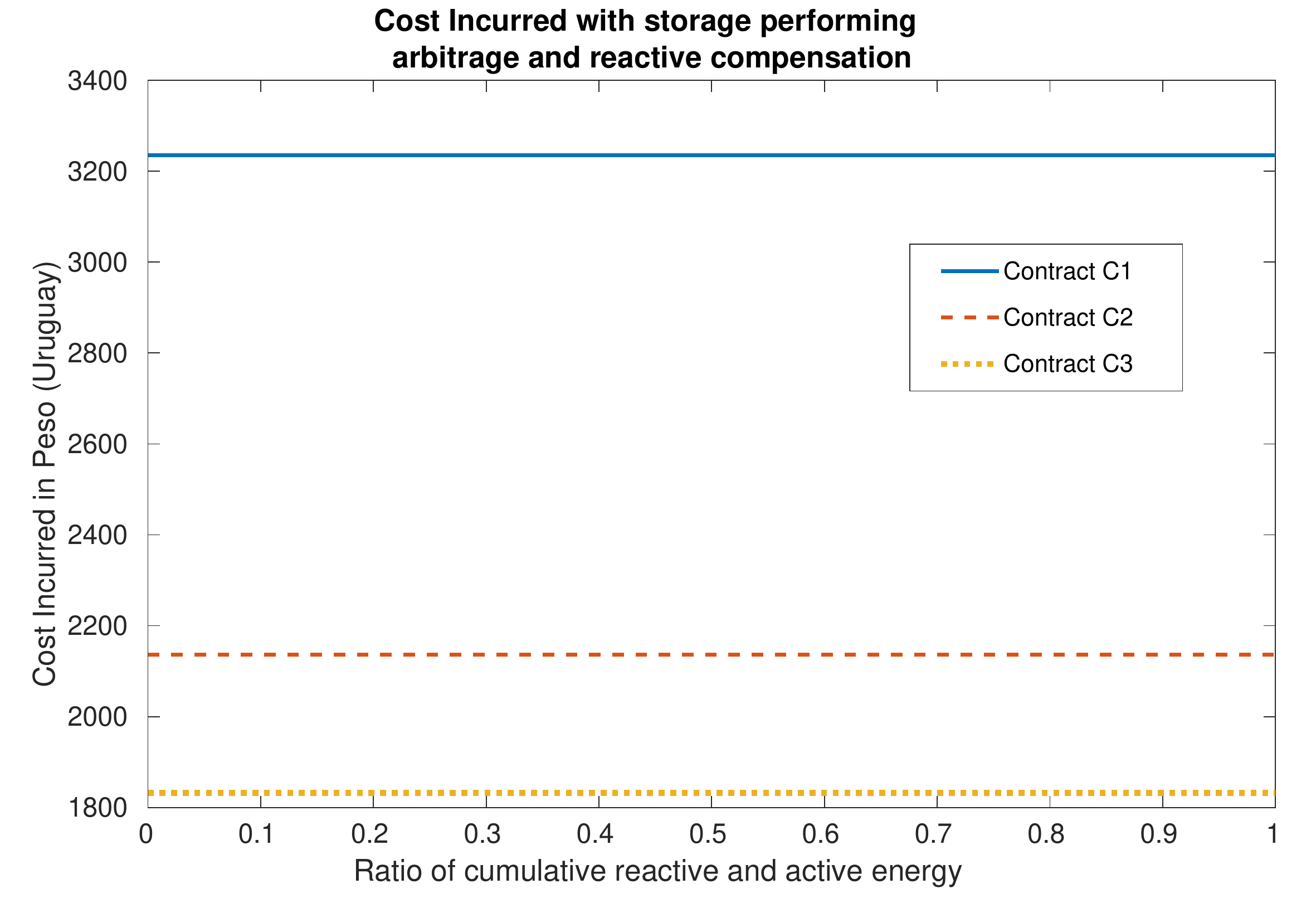}
	\caption{Total cost of electricity for Tesla PowerWall 2}\label{res12}
\end{figure}

\begin{figure}
	\centering
	\includegraphics[width=3.2in]{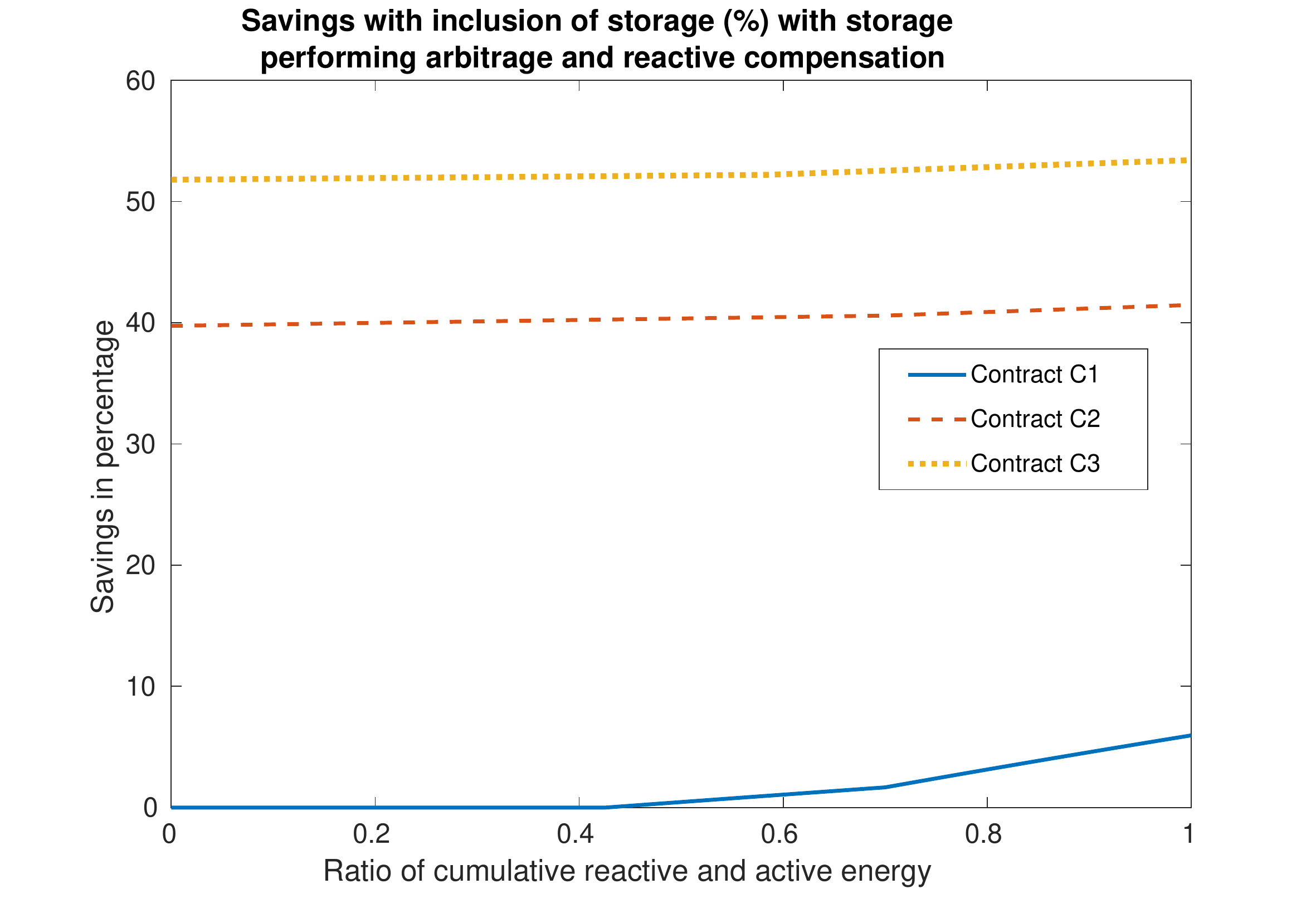}
	\caption{Profit due to inclusion of storage in terms of savings in percentage for Tesla PowerWall 2}\label{res13}
\end{figure}

Fig.~\ref{res12} presents the total cost of consumption (includes fixed and power cost) with energy storage performing arbitrage and reactive power compensation. Clearly, contract C3 consumer pays the least and consumer under C1 pays the maximum electricity bill in this scenario with Tesla PowerWall 2.
Fig.~\ref{res13} shows the percentage of consumer would make with respect to the nominal case under the same contract. Consumers with C3 would make a profit exceeding 50\% by installing Tesla PowerWall 2. 
Consumers with C2 would make a profit exceeding 40\% by installing Tesla PowerWall 2. 
Note since consumer under C1 could perform only reactive power compensation, it makes a profit only if the ratio of reactive energy and active energy exceeds 0.426.

We observe that consumers with storage should opt for C3, i.e., three level time-of-use electricity pricing rather than fixed pricing under C1. Note that the selection of contract also depends on the reactive energy consumed, load and storage size. In this analysis we assume that conditions $T_{\text{off-peak}} > T_{\text{ch}}$ and $T_{\text{peak}} > T_{\text{dis}}$ are true.

\subsection{Energy Storage Profitability}
\label{profitabilitystorage}
The cost of Tesla Powerwall 1 is \$3000 and Tesla Powerwall 2 is \$5500, making the per kWh cost of \$470 and \$398 \cite{tesla}. Consider the storage could perform 3000 cycles of 100\% depth-of-discharge over a maximum calendar life of 10 years. We use the degradation model presented in \cite{hashmi2018long}.
In order to have the storage application profitable the per cycle gains should satisfy\\
$\bullet$ For Tesla Powerwall 1: the battery should earn more than \$1 for each cycle in order to be profitable,\\
$\bullet$ For Tesla Powerwall 2: the battery should earn more than \$1.8333 for each cycle in order to be profitable.

The depth-of-discharge of battery for each day is equivalent to $\text{DoD}_{\text{daily}} = \text{SoC}_{\max} - \text{SoC}_{\min}$. Thus, the battery performs 0.7608 cycles of 100\% DoD each day and 277.7 cycles each day.

		\begin{table*}
			\centering
			\begin{center}
			\caption {\small{Profitability}}
			\label{gainspercycle}
				\begin{tabular}{| c | c| c|}
					\hline
					Parameter& Tesla Powerwall 1 & Tesla Powerwall 2\\ 
					\hline
					\hline
					Desired \$/cycle & 	\$1.0 & \$1.833\\
					Month gain in pesos& C2: 682.5, C3:942.6 & C2: 1439.6, C3:1988.3\\
					Month gain in dollars& C2: 20.29, C3:28.02 &C2: 42.8, C3:59.11\\
					Total cycles in month & 22.83& 22.83\\
					Gain in \$/cycle &	\textcolor{red}{C2: 0.889}, C3: 1.227  & C2: 1.874, \textbf{C3: 2.589} \\
					\hline
				\end{tabular}
				\hfill\
			\end{center}
		\end{table*}

Of course the storage gains are dependent on the load active and reactive energy consumed during different times of the day. For the nominal load profile storage gains by only performing arbitrage appears to be profitable as the dollar per cycle gain exceeds the dollars per cycle cost calculated based on the cost and life ratings of the battery. Adding the reactive compensation gains would further improve the dollars per cycle of the battery, in effect reducing the payback period.
The simple payback period for cases where storage is profitable would lie between 3-4 years, similar to storage returns in the island of Madeira \cite{hashmi2019energy}.

\section{Conclusion}
\label{sectionUrug7}
The details of electricity billing and net-metering policy under the new electricity consumer contracts applicable to the low voltage electricity consumers in Uruguay are summarized. The consumer contracts make it attractive for consumers to install energy storage and distributed generation.
Due to the huge difference between peak and off-peak electricity pricing, installing energy storage could be rewarding for LV consumers. Furthermore, consumers could use the storage converter for reactive power compensation which could provide an additional revenue stream for the consumer. For each contract threshold-based uncertainty insensitive algorithm based on hierarchical control of active energy for arbitrage and reactive energy compensation is proposed.
In the present work, the question of whether energy storage is profitable
for three new energy contracts in Uruguay is assessed. These contracts are
of particular interest for two reasons. Firstly, the buying price and selling
price are the same, for which it is shown that policy for performing arbitrage
is insensitive to load and generation. Secondly, reactive power compensation is
monetarily compensated in some of those cases. With data from real batteries and taking into account battery degradation,
it is shown that for some of the new contracts it will be profitable to buy a battery,
while for others it will not. For the latter case, an evaluation of the best contract
available is also provided.
Using numerical results we show that arbitrage is more
profitable that reactive energy compensation, but prior work \cite{hashmi2018pfc} identifies that reactive compensation can still be performed
without compromising the former.
Energy markets with similar pricing and net-metering policy could use this analysis.

	\bibliographystyle{model2-names}
	
	\bibliography{uruguay_ref}

	\appendix
\section{Arbitrage with Net-Energy Metering with Sell Price equal to Retail Rate}
\label{arbitragesec}
Energy arbitrage refers to buying energy when the price of electricity is low and selling it when price is high. Another interpretation could be shifting consumption from high price periods to low price periods. The problem of energy arbitrage is considered from consumer or end-user perspective. The consumer of electricity has its goal to minimize the overall cost of consumption by installing energy storage device such as a battery for performing energy arbitrage. 
The price of electricity is denoted as ${p}_{elec}(i)$. The consumer optimization problem is given as \cite{hashmi2017optimal} \vspace{-10pt}
\begin{gather*}
\text{($P_{arbitrage}^0$)} \quad
\min_{P_B^i \in \left[\eta_{\text{dis}}\delta_{\min}h , \frac{\delta_{\max}h}{\eta_{\text{ch}}}\right]}  \sum_{i=1}^N [P_i + P_B^i] {p}_{elec}(i) \\
\text{subject to,} 
\text{(i.) } 
b_{\min} - b_0\leq \sum_{j=1}^i x_j \leq b_{\max}- b_0.
\end{gather*}
The optimization problem ($P_{arbitrage}^0$) is equivalent to
$ \min$ $\sum_{i=1}^N \{ {p}_{elec}(i) P_i + {p}_{elec}(i) P_B^i\}$.
Since there is no degree of freedom in ${p}_{elec}(i)P_i$ for all $i$, therefore, ($P_{arbitrage}^0$) is equivalent to
\begin{gather*}
\text{($P_{arbitrage}^{equi}$)} \quad
\min_{P_B^i \in \left[\eta_{\text{dis}}\delta_{\min}h , \frac{\delta_{\max}h}{\eta_{\text{ch}}}\right]}  \sum_{i=1}^N P_B^i {p}_{elec}(i) \\
\text{s.t. } 
\text{(i.) } 
b_{\min} - b_0\leq \sum_{j=1}^i x_j \leq b_{\max}- b_0 , \forall i \in \{1,..,N\}
\end{gather*}
%

\section{Proof of Theorem~\ref{profitabilityarb}}
\label{theorem41}
\begin{proof}
	Let $\epsilon$ denote a sufficiently small amount of energy, and w.l.o.g assume that inside every time slot, all energy packets arrive sequentially. We define $E$=$\{\epsilon_1, \epsilon_2,.., \epsilon_K  \}$ as a complete order of all energy packets that charged the battery. Let $I_k$ denote the time slot at which the packet $\epsilon_k$ entered the battery and $O_k$ when it left it. If the packet never left (that is, if there is remaining energy in the battery at the end of the arbitrage) we let $k=\infty$ and $p^\infty_s = 0$.
	Let $n_{ij}$ denote the number of packets that charged the battery during time slot $i$ and discharged at time slot $j$.  
	\begin{gather*}
	n_{ij} = \sum_{k=0}^K \mathbb{I}\{ I_k = i , O_k = j\}
	\end{gather*}
	The price of charging the battery $\epsilon$ at time $i$ is given by $\epsilon\frac{ p^i_b}{\eta_{ch}}$. The earnings of discharging the battery at time slot $j$ is given by $\epsilon \eta_{dis} p^j_s$.
	Finally, the total profit obtained by the arbitrage is given by:
	\begin{gather*}
	\text{Profit} =  \sum_{i\in[N], j\in[N]\backslash\{1\} \cup \{\infty\}} \epsilon n_{ij} \left[\eta_{dis} p^j_s -  \frac{ p^i_b}{\eta_{ch}}\right]
	\end{gather*}
	Because $m_{ij} \in \mathbb{N}$, if $p^j_s \eta_{dis} \leq \frac{p^i_b}{\eta_{ch}}$, then $Profit \leq 0$.	
\end{proof}
\begin{figure}
	\centering
	\includestandalone[mode=buildnew, width=2.8in]{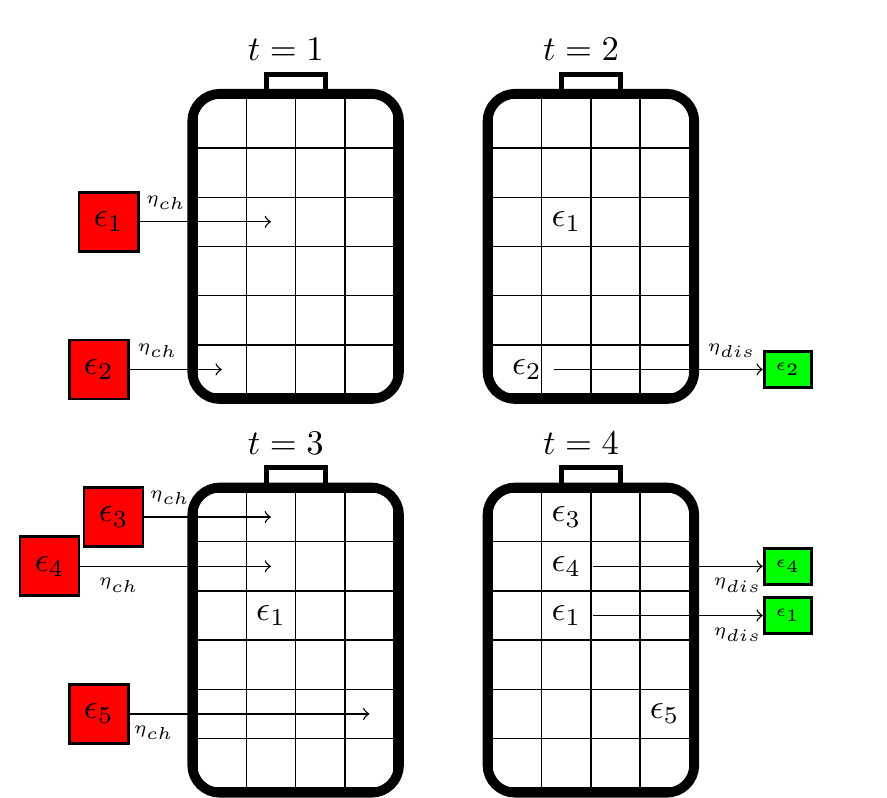}
	\caption{The size of the squares represents the amount of energy. Red squares (before been charged in the battery) are bigger than the squares composing the battery. This in turn, are bigger than the green squares that output the battery. Regarding the case presented: initially, the battery is empty. At $t=1$, 2 energy packets arrive $\epsilon_1$ and $\epsilon_2$, so $I_1 = I_2 = 1$. In the next timeslot, $\epsilon_2$ leaves so we can update $O_2 = 2$. At timeslot $t=3$, three energy packets arrive (3,4,5) so we have $I_3 = I_4 = I_5 = 3$. Finally, in the last timeslot the energy packets $\epsilon_1$ and $\epsilon_4$ leave so $O_1 = O_4 = 4$. Because the battery is still fully charged at the end of the horizon, $O_3 = O_5 = \infty$. In this example $n_{12} = 1$, $n_{14} = 1$, $n_{34} = 1$ and $n_{3\infty } = 2$. With this information the final profit can be found.}
	\label{fig:energypackets}
\end{figure}

\end{document}